\documentclass[aps,prb,reprint,superscriptaddress]{revtex4-2}

\usepackage{amsmath}
\usepackage{upgreek}
\usepackage{float}
\usepackage{graphicx}
\usepackage{tabularx}
\usepackage{hyperref}

\begin{document}

\title{A Hybrid Density Functional Theory Benchmark Study on Lithium Manganese Oxides}

\author{Marco Eckhoff}
\email{marco.eckhoff@chemie.uni-goettingen.de}
\affiliation{Universit\"at G\"ottingen, Institut f\"ur Physikalische Chemie, Theoretische Chemie, Tammannstra{\ss}e 6, 37077 G\"ottingen, Germany.}
\author{Peter E. Bl\"ochl}
\email{peter.bloechl@tu-clausthal.de}
\affiliation{Technische Universit\"at Clausthal, Institut f\"ur Theoretische Physik, Leibnizstra{\ss}e 10, 38678 Clausthal-Zellerfeld, Germany.}
\affiliation{Universit\"at G\"ottingen, Institut f\"ur Theoretische Physik, Friedrich-Hund-Platz 1, 37077 G\"ottingen, Germany.}
\author{J\"org Behler}
\email{joerg.behler@uni-goettingen.de}
\affiliation{Universit\"at G\"ottingen, Institut f\"ur Physikalische Chemie, Theoretische Chemie, Tammannstra{\ss}e 6, 37077 G\"ottingen, Germany.}
\affiliation{Universit\"at G\"ottingen, International Center for Advanced Studies of Energy Conversion (ICASEC), Tammannstra{\ss}e 6, 37077 G\"ottingen, Germany}

\date{\today}

\begin{abstract}
The lithium manganese oxide spinel Li$_x$Mn$_2$O$_4$, with $0\leq x\leq 2$, is an important example for cathode materials in lithium ion batteries. However, an accurate description of Li$_x$Mn$_2$O$_4$ by first-principles methods like density functional theory is far from trivial due to its complex electronic structure, with a variety of energetically close electronic and magnetic states. It was found that the local density approximation as well as the generalized gradient approximation (GGA) are unable to describe Li$_x$Mn$_2$O$_4$ correctly. Here, we report an extensive benchmark for different Li$_x$Mn$_y$O$_z$ systems using the hybrid functionals PBE0 and HSE06, as well as the recently introduced local hybrid functional PBE0r. We find that all of these functionals yield energetic, structural, electronic, and magnetic properties in good agreement with experimental data. The notable benefit of the PBE0r functional, which relies on on-site Hartree-Fock exchange only, is a much reduced computational effort that is comparable to GGA functionals. Furthermore, the Hartree-Fock mixing factors in PBE0r are smaller than in PBE0, which improves the results for (lithium) manganese oxides. The investigation of Li$_x$Mn$_2$O$_4$ shows that two Mn oxidation states, +III and +IV, coexist. The Mn$^\text{III}$ ions are in the high-spin state and the corresponding MnO$_6$ octahedra are Jahn-Teller distorted. The ratio between Mn$^\text{III}$ and Mn$^\text{IV}$ and thus the electronic structure changes with the Li content while no major structural changes occur in the range from $x=0$ to $1$. This work demonstrates that the PBE0r functional provides an equally accurate and efficient description of the investigated Li$_x$Mn$_y$O$_z$ systems.
\end{abstract}


\maketitle

\section{Introduction}\label{sec_introduction}

Today, a world without lithium ion batteries is hard to imagine because they are essential for the energy supply of almost all portable electronic devices from mobile phones to laptop computers. The lithium manganese oxide spinel Li$_x$Mn$_2$O$_4$, an intercalation compound with Li contents $0\leq x\leq 2$, is a prominent example for cathode materials in lithium ion batteries,\cite{Thackeray1997} which offers advantages such as low costs and non-toxicity.\cite{Berg1999}
\\
\begin{figure}[t!]
\centering
\includegraphics[width=\columnwidth]{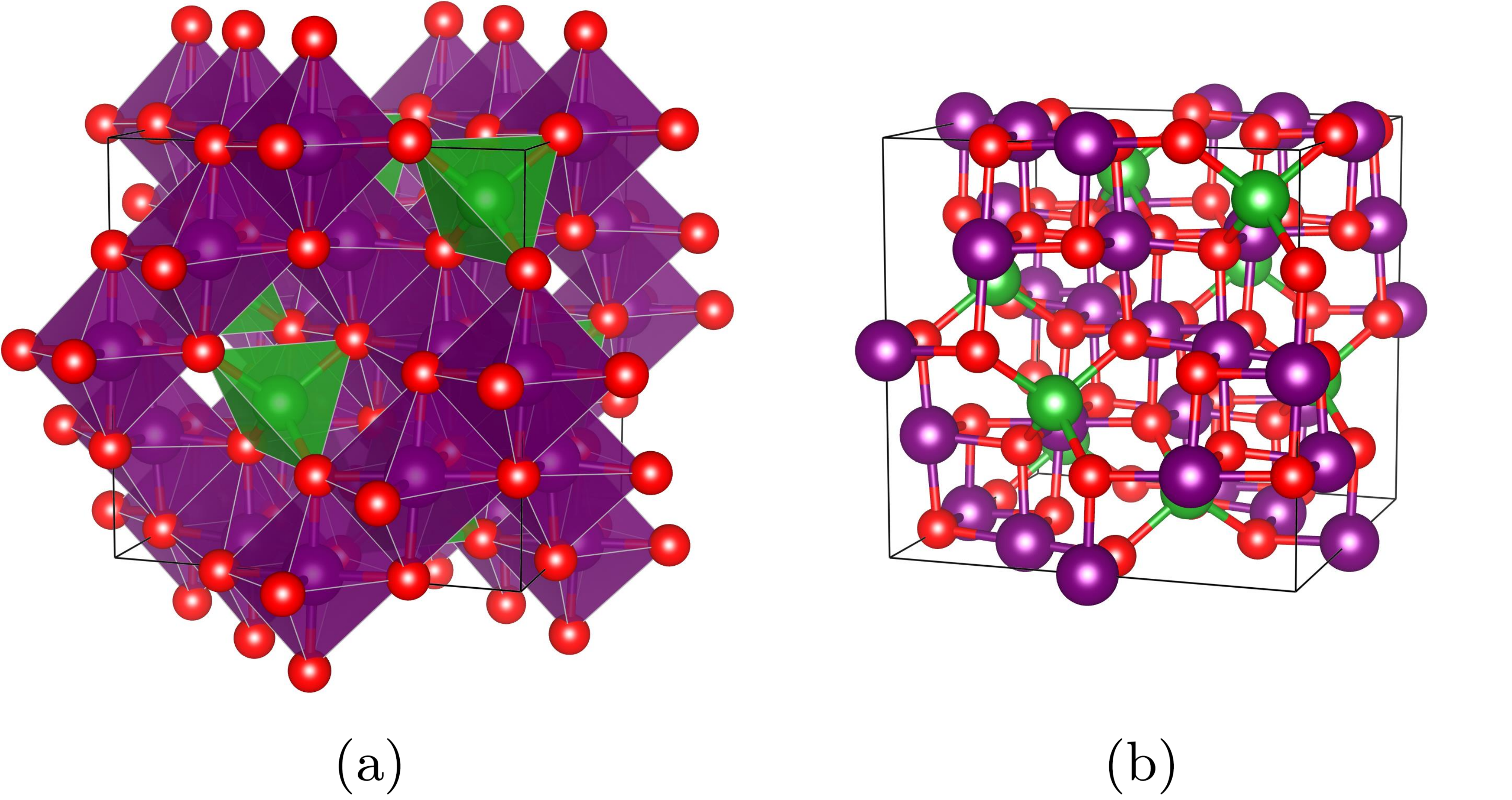}
\caption{Spinel structure of LiMn$_2$O$_4$.\cite{Akimoto2004} Li is colored green, O red, and Mn purple. The unit cell is marked by black lines. Panel (a) shows the coordination polyhedra of Li (tetrahedra) and Mn (octahedra) while (b) includes only the atoms of the unit cell containing eight formula units. This and all other figures in this work were created with VESTA version 3.4.4.\cite{Momma2011}}\label{fig_limn2o4_structure}
\end{figure}
The Li content can be varied over a wide range of $0\leq x\leq 2$ using, for example, electrochemical (de)intercalation.\cite{Tarascon1994} At temperatures above\ $\sim$\,290\,K, for compositions $0\leq x\leq 1$ the crystal exhibits a cubic spinel structure with the space group Fd$\overline{3}$m in which Li occupies the tetrahedral 8a sites and Mn the octahedral 16d sites (FIG.\ \ref{fig_limn2o4_structure}).\cite{Akimoto2004} The MnO$_6$ octahedra share one half of their edges with other MnO$_6$ octahedra. Each of their corners is shared with one LiO$_4$ tetrahedron in the case of LiMn$_2$O$_4$. The MnO$_6$ octahedra build a superstructure of corner-sharing $\left(\text{MnO}_6\right)_4$ tetrahedra. Complete delithiation results in the formation of $\uplambda$-Mn$_2$O$_4$ exhibiting the same manganese oxide host lattice, while all tetrahedral sites are unoccupied. Thus, no major internal structural changes occur when reducing the Li contents from $x=1$ to $x=0$, while the lattice constant is approximately linearly decreasing. The varying content of Li ions is balanced by changes in the oxidation states of the Mn ions to maintain overall charge neutrality. Consequently, in LiMn$_2$O$_4$, one half of the Mn ions is in the high-spin (hs) Mn$^\text{III}$ state (t$_\mathrm{2g}^3$e$_\mathrm{g}^1$) while the other half is in the oxidation state Mn$^\text{IV}$ (t$_\mathrm{2g}^3$e$_\mathrm{g}^0$).\cite{Berg1999} The lithium-free $\uplambda$-Mn$_2$O$_4$ contains exclusively Mn$^\text{IV}$ ions.
\\
The hs-Mn$^\text{III}$ ions lead to Jahn-Teller (JT) distorted MnO$_6$ octahedra.\cite{Jahn1937} Therefore, the cubic crystal structure can be understood as a disordered arrangement of Mn$^\text{III}$ and Mn$^\text{IV}$,\cite{Ouyang2009} in which the JT distortions are thermally averaged in all spatial directions. Below\ $\sim$\,290\,K, an increasing ordering of the Mn ions and the alignment of the distorted octahedra results in a transformation of the spinel structure to an orthorhombic phase with space group Fddd.\cite{Piszora2004} At very low temperatures, an antiferromagnetic long-range order as well as spin-glass behavior were observed.\cite{Tomeno2001, Sugiyama1997}
\\
For $x>1$, a phase transition takes place to a tetragonal spinel structure with space group I4$_1$/amd.\cite{Berg1999} This transition is caused by the increasing number of Mn$^\text{III}$ ions and the associated JT distorted MnO$_6$ octahedra upon lithiation. A miscibility gap exists between the cubic ($x=1$) and tetragonal ($x=2$) spinel structure.\cite{Maier2016} If the average Li content is higher than $x=1$, the tetragonal phase with $x=2$ will form and both phases coexist in the range $1<x<2$.
\\
Detailed insights into the structure of Li$_x$Mn$_2$O$_4$ have been gained in various experiments, e.g.\ X-ray diffraction, transmission electron microscopy, and atom probe tomography.\cite{Koehler2017, Maier2016} Theoretical studies are able to provide complementary information, for example, about the underlying atomistic processes of the phase transitions or about the Li diffusion pathways. However, the underlying electronic structure of Li$_x$Mn$_2$O$_4$ is very complex due to the large number of energetically close electronic and magnetic states. Therefore, a theoretical treatment using density functional theory (DFT) is far from trivial. Previous studies could show that the local density approximation (LDA) as well as the generalized gradient approximation (GGA) do not yield a qualitatively correct electronic structure,\cite{Ouyang2009, Mishra1999, vanderVen2000, Grechnev2002} since all Mn ions in LiMn$_2$O$_4$ are found to be in an averaged oxidation state of 3.5. Further, in contrast to experiment no band gap exists. Consequently, GGA$+U$ or hybrid functionals are required to correctly obtain distinct Mn$^\text{III}$ and Mn$^\text{IV}$ ions as well as a qualitatively correct band gap,\cite{Ouyang2009, Chevrier2010} which is about 1.2\,eV in experiment.\cite{Ouyang2006}
\\
Manganese is known for its wide range of possible oxidation states and complex magnetic structures of its compounds. Experimental studies could show that even elemental $\upalpha$-Mn exhibits a non-collinear antiferromagnetic structure.\cite{Hobbs2001} The Mn$^\text{II}$ ions in MnO are arranged in a way that the antiferromagnetic order is present in all three cubic directions.\cite{Franchini2007} The magnetic structure of the Mn$^\text{III}$ ions in $\upalpha$-Mn$_2$O$_3$ was investigated by a combined GGA$+U$ and neutron diffraction study.\cite{Cockayne2013} The outcome is a complex non-collinear antiferromagnetic order. In $\upbeta$-MnO$_2$ the Mn$^\text{IV}$ ions crystallise in an antiferromagnetic structure with helically ordered magnetic moments.\cite{Yoshimori1959}
\\
\begin{figure}[t!]
\centering
\includegraphics[width=\columnwidth]{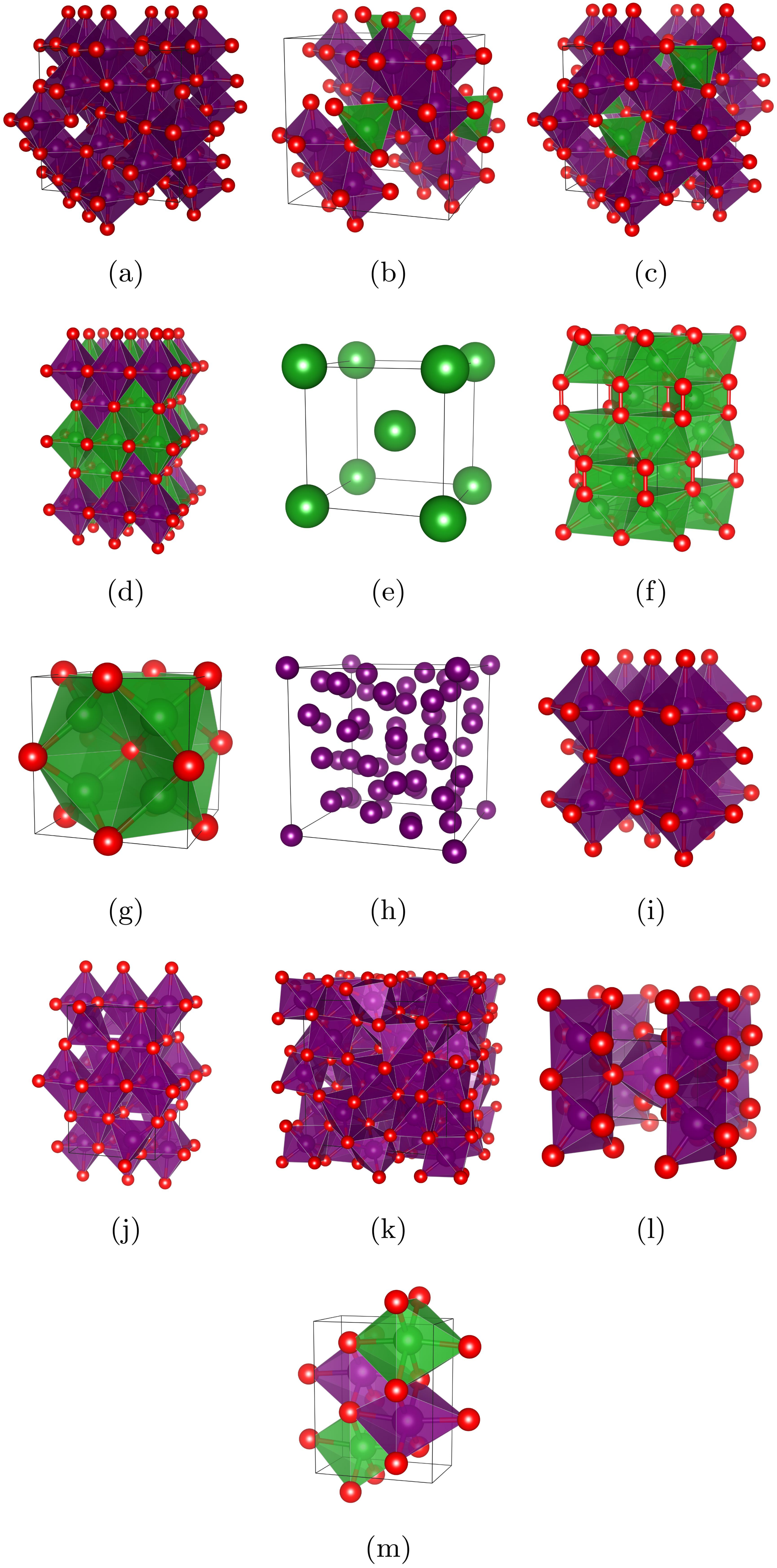}
\caption{Structures of (a): $\uplambda$-Mn$_2$O$_4$,\cite{Takahashi2003} (b): Li$_{0.5}$Mn$_2$O$_4$,\cite{Bianchini2015} (c): LiMn$_2$O$_4$,\cite{Akimoto2004} (d): Li$_2$Mn$_2$O$_4$,\cite{Mosbah1983} (e): Li,\cite{Wyckoff1963} (f): Li$_2$O$_2$,\cite{Foeppl1957} (g): Li$_2$O,\cite{David2007} (h): $\upalpha$-Mn,\cite{Wyckoff1963} (i): MnO,\cite{Taylor1984} (j): Mn$_3$O$_4$,\cite{Jarosch1987} (k): $\upalpha$-Mn$_2$O$_3$,\cite{Geller1971} (l): $\upbeta$-MnO$_2$,\cite{Bolzan1993} and (m): LiMnO$_2$.\cite{Kellerman2007} Li is colored green, O red, and Mn purple. The black lines represent the unit cell.\cite{Momma2011} }\label{fig_structures}
\end{figure}
In this study, we investigate the accuracy of the well-established PBE0\cite{Perdew1996a, Adamo1999} and HSE06\cite{Heyd2003, Heyd2006, Krukau2006} functionals as well as the recently developed local hybrid functional PBE0r\cite{Sotoudeh2017} for these structures and a variety of other systems containing lithium, manganese, and oxygen. The benchmarked properties include formation energies, structural properties, and the density of states. Furthermore, the magnetic order and the intercalation potential of Li$_x$Mn$_2$O$_4$ are determined. With this extensive and rigorous benchmark we examine the quality of hybrid DFT calculations for this class of materials and test the approximations in HSE06 and PBE0r compared to PBE0 to increase the efficiency. Especially, we optimize the admixture of Hartree-Fock (HF) exchange in the PBE0r hybrid functional for the calculation of Li$_x$Mn$_y$O$_z$ systems. We validate its approach of including only on-site HF exchange terms and keep the off-site terms on the GGA level aiming a very efficient functional which provides a high accuracy for these systems. Additionally, in the Supporting Information (SI) the D3 method\cite{Grimme2010} is evaluated for these systems as a possible approximate correction to overcome limitations in the description of van der Waals interactions in current hybrid functionals.
\\
The main focus of our work is on the Li$_x$Mn$_2$O$_4$ spinel structure. Therefore, our benchmark set includes spinels with varying Li contents, specifically the X-ray diffraction structures of $\uplambda$-Mn$_2$O$_4$ (at room temperature),\cite{Takahashi2003} Li$_{0.5}$Mn$_2$O$_4$ (at 293\,K),\cite{Bianchini2015} LiMn$_2$O$_4$ (at 330\,K),\cite{Akimoto2004} and Li$_2$Mn$_2$O$_4$ (at room temperature)\cite{Mosbah1983} (FIG.\ \ref{fig_structures} (a)-(d)). Moreover, several related systems, which were thoroughly investigated theoretically and experimentally before, were chosen to benchmark the exchange-correlation functionals. They comprise the X-ray diffraction structures of Li (at 78\,K),\cite{Wyckoff1963} Li$_2$O$_2$,\cite{Foeppl1957}, Li$_2$O,\cite{David2007} $\upalpha$-Mn,\cite{Wyckoff1963} MnO,\cite{Taylor1984} Mn$_3$O$_4$,\cite{Jarosch1987} $\upalpha$-Mn$_2$O$_3$,\cite{Geller1971} $\upbeta$-MnO$_2$ (neutron diffraction),\cite{Bolzan1993} and orthorhombic LiMnO$_2$\cite{Kellerman2007} (FIG.\ \ref{fig_structures} (e)-(m)). These structures, which refer to room temperature unless stated differently, cover both common Li oxidation states of 0 and +I and the Mn oxidation states 0, +II, +III, and +IV. This benchmark set enables us to find optimal settings for the PBE0r functional to describe most of the Li$_x$Mn$_y$O$_z$ systems accurately. Furthermore, the molecules H$_2$, O$_2$, and H$_2$O are part of the benchmark set because O$_2$ is required as reference for the calculation of the formation energies and its total energy is checked using the formation energy of water. Moreover, we use a water molecule as an example for a covalently bonded system in the discussion of the results.

\section{Methods}

While Kohn-Sham DFT is in principle exact, approximate exchange-correlation functionals $E_{\mathrm{xc}}$ need to be employed. In recent decades, a hierarchy of functionals was proposed, which can have a notable impact on the quality of the obtained results.\cite{P3910,P3883,P4009} In this work, we will address the performance of modern hybrid functionals, which currently represent the state-of-the-art, using the examples PBE0, HSE06, and PBE0r.
\\
PBE0\cite{Perdew1996a, Adamo1999} is based on the PBE\cite{Perdew1996} GGA functional, but 25{\%} of the PBE exchange $E_\mathrm{x}^\mathrm{PBE}$ are replaced by exact HF exchange $E_\mathrm{x}^\mathrm{HF}$ resulting in
\begin{align}
E_\mathrm{xc}^\mathrm{PBE0}=E_\mathrm{xc}^\mathrm{PBE}+\dfrac{1}{4}\left(E_\mathrm{x}^\mathrm{HF}-E_\mathrm{x}^\mathrm{PBE}\right)\ .
\end{align}
The calculation of the HF exchange increases the computational costs drastically due to the long-range nature of the Coulomb interaction. Replacing the Coulomb interaction in the exchange by a screened interaction reduces the number of integrals to be evaluated, and it recovers the subtle balance of exchange and correlation for electrons at large distances. In the HSE06 functional \cite{Heyd2003, Heyd2006, Krukau2006} this is realized by a screened Coulomb potential,
\begin{align}
\dfrac{1}{r}=\dfrac{1-\mathrm{erf}(\omega r)}{r}+\dfrac{\mathrm{erf}(\omega r)}{r}\ ,
\end{align}
i.e.\ the Coulomb potential is separated into a short-range and a long-range part. The separation range is determined by the screening parameter, $\omega=0.11\,a_0$, which was empirically obtained by calibration to experimental properties.\cite{Krukau2006} $a_0$ is the Bohr radius. Consequently, the HF exchange is only calculated for the short-range part (superscript s) but not for the long-range part (superscript l), which reduces the computational cost substantially for extended systems,
\begin{align}
E_\mathrm{xc}^\mathrm{HSE06}=\dfrac{1}{4}E_\mathrm{x}^\mathrm{HF,\,s}+\dfrac{3}{4}E_\mathrm{x}^\mathrm{PBE,\,s}+E_\mathrm{x}^\mathrm{PBE,\,l}+E_\mathrm{c}^\mathrm{PBE}\ .
\end{align}
The correlation part $E_\mathrm{c}^\mathrm{PBE}$ is not affected by the screened Coulomb potential. For a screening parameter $\omega=0$ the functional is equal to PBE0. For $\omega=\infty$ the functional is identical to PBE.
\\
The recently published PBE0r functional\cite{Sotoudeh2017} is a local hybrid exchange-correlation functional which is also derived from PBE0. The Kohn-Sham orbitals are mapped onto a minimal basis of localized atom-centered tight-binding orbitals. The tight-binding orbitals are used to calculate the on-site HF exchange terms including the exchange interaction between core and valence electrons. All other exchange contributions, i.e.\ those with tight-binding orbitals centered on different atoms, are neglected. Hence, PBE0r can be regarded as range-separated hybrid functional where the cutoff of the exchange interaction is defined by the localized tight-binding orbitals. The PBE0r exchange-correlation functional $E_\mathrm{xc}^\mathrm{PBE0r}$ is given by
\begin{align}
E_\mathrm{xc}^\mathrm{PBE0r}=E_\mathrm{xc}^\mathrm{PBE}+\sum_{n=1}^{N}a_n\left(E_{\mathrm{x},\,n}^\mathrm{HF,\,r}-E_{\mathrm{x},\,n}^\mathrm{PBE,\,r}\right)\ .
\end{align}
Since the inclusion of HF exchange is restricted to the on-site terms, $E_\mathrm{x}^\mathrm{HF,\,r}$, only the corresponding PBE exchange terms, $E_\mathrm{x}^\mathrm{PBE,\,r}$, are subtracted in order to avoid double counting. The HF mixing factor $a_n$ of the $N$ atoms can vary for the chemical elements in a given system. For the determination of the HF mixing factors different routes were applied in previous studies: fitting according to ground-state properties,\cite{He2012} using the inverse of the dielectric constant,\cite{Alkauskas2010, Marques2011} employing a dielectric model dependence\cite{Chen2018, Cui2018, Liu2020}, applying self-consistent schemes,\cite{Skone2014, He2017} or derivation from perturbation theory arguments as it was the case for PBE0.\cite{Perdew1996a} We choose to perform a systematic search on a grid of HF mixing factors which can be different for each element to get a set of mixing factors which yields good agreement with known experimental and theoretical reference data. Using the grid approach we can identify trends which lead us to a good compromise for all reference data. The procedure of the search and the trends are described in the SI.
\\
Our empirically determined optimal mixing factors for the Li$_x$Mn$_y$O$_z$ systems are between 0.05 and 0.09. Lower mixing factors than the PBE0 value of 0.25 are also applied in previous studies on transition-metal perovskites including Mn, which employ mixing factors of 0.15 in HSE06\cite{He2012} and between 0.07 and 0.15 in PBE0r.\cite{Sotoudeh2017} Moreover, a work on transition-metal complexes concludes that the optimal admixture of HF exchange is between 0.08 and 0.16 in B3LYP\cite{Becke1993} for the calculation of Fe$^\mathrm{II}$-S complexes.\cite{Reiher2001}
\\
Replacing the local exchange used in LDA and GGA functionals by exact exchange terms of HF has a strong affect on the Kohn-Sham band structure. Note, that the GW method\cite{Hedin1965}, a many-body Green's functional method, has a similar structure as the HF method, albeit with a screened interaction in the exchange term instead of the long-range Coulomb interaction. Therefore, it is not surprising that the Kohn-Sham band structure of hybrid functionals tend to agree better with quasiparticle spectra than that of local density functionals such as LDA and GGA. In order to rationalize different types of hybrid functionals, we find it useful to distinguish three types of HF exchange terms:
\\
1.\,\,On-site exchange acts between orbitals centered on the same site. The main effect can be attributed to the self-interaction correction: Otherwise degenerate states split into a multiplet of filled and another one of empty orbitals. The two bands can be attributed to the Mott-Hubbard bands and their separation is roughly proportional to the $U$-parameter. On-site exchange is important in narrow, partially filled d and f shells as they are present in many transition metal oxides.
\\
2.\,\,Bond exchange consists of an exchange term, for which a density on one site interacts with that of another site. This exchange term is sensitive to the phase relation of the orbitals on the two sites and it distinguishes bonding and antibonding states. Bond exchange opens the band gap of covalent materials such as silicon.
\\
3.\,\,Long-range exchange is analogous to bond exchange. It acts over longer distances than bond distances. This term affects metallic solids and results for a free electron gas in a vanishing density of states at the Fermi level. This behavior is caused by the long-range tail of the Coulomb interaction in the exchange terms, which is effectively removed by screening.
\\
GGAs such as PBE describe many transition metal oxides poorly, because they lack exact on-site exchange which splits the d shell into filled and empty orbitals. The on-site exchange terms are, however, well captured by the PBE0r functional. On the other extreme of local functionals, the PBE0 functional overestimates long-range exchange, so that solids, in particular metals, are not described adequately.\cite{Marsman2008} This problem is remedied by range separated hybrid functionals such as HSE06. While one of the motivations for HSE06 was to limit the computational effort, it also captures an important physical effect, namely screening. In the PBE0r functional the range separation is carried further: it removes not only long-range exchange but also bond exchange. Except for the replaced on-site exchange terms PBE0r is the same as PBE, i.e.\ PBE0r does not miss any terms. Due to the drastically reduced number of included HF terms compared to PBE0, whose calculation would account for most of the computational costs, the computational effort of PBE0r is comparable to GGA functionals. While PBE0r is inadequate to describe covalent materials such as silicon on a higher level than PBE, it is suitable for an accurate description of the Li$_x$Mn$_y$O$_z$ systems discussed in the present study.

\section{Computational Details}

The PBE0r calculations were performed using the Car-Parrinello Projector Augmented-Wave (CP-PAW) code (version from 28$^\text{th}$ September 2016) which applies the projector augmented-wave (PAW) method\cite{Bloechl1994} for electronic structure calculations. The augmentation of the PAW method included the 1s orbital of H, the 2s and 2p orbitals of Li, the 2s, 2p, and 3d orbitals of O, and the 3s, 3p, 3d, 4s, and 4p orbitals of Mn. In the case of Mn, besides the 4s and 3d orbitals also the 3s and 3p orbitals were treated as valence electrons, because these semi-core states are required to describe the electronic structure of Mn in the systems of the benchmark set properly. The matching radii for the construction of the auxiliary partial waves in units of the covalent radii were set to 0.7 for all orbitals. The covalent radii were set to 0.32\,{\AA} for H, 1.23\,{\AA} for Li, 0.73\,{\AA} for O, and 1.17\,{\AA} for Mn. The auxiliary wave functions were constructed as node-less partial waves.\cite{Bloechl2012} The tight-binding orbitals include the 1s orbital of H, the 2s orbital of Li, the 2s and 2p orbitals of O, and the 3s, 3p, 3d, and 4s orbitals of Mn. The mixing factors $a_m$ for the HF exchange were adjusted to minimize the errors of the formation energies and band gaps of the benchmark set using experimental reference data, which are given and referenced in the following chapter. The obtained $a_m$ values are 0.07 for H, 0.07 for Li, 0.05 for O, and 0.09 for Mn. The determination of the given mixing factors is described in detail in the SI. Moreover, the complete settings for each element are given in the SI.
\\
As the non-collinear treatment of the spins would increase the computational effort for the benchmark systems significantly, the approximation of collinear spin-polarization was applied. The plane wave cutoff was 25\,$E_\mathrm{H}$ (Hartree) for the auxiliary wave functions and 100\,$E_\mathrm{H}$ for the auxiliary densities. With these settings, the obtained formation energies deviate less than 0.01\,eV per atom from the complete basis set limit. The $\boldsymbol{\Gamma}$-centered $\mathbf{k}$-point grid was set to $2\times2\times2$ for the LiMn$_2$O$_4$ unit cell, and for the other systems $\mathbf{k}$-point grids of a comparable $\mathbf{k}$-point density were chosen. For metallic systems, the improved tetrahedron method was used.\cite{Bloechl1994a} This ensures a convergence level of about 0.001\,eV per atom for energies differences. Molecular systems were placed in a large periodic cell with lattice vectors (0\ 11.5\ 11.5)$^\mathsf{T}$\,{\AA}, (12\ 0\ 12)$^\mathsf{T}$\,{\AA}, and (12.5\ 12.5\ 0)$^\mathsf{T}$\,{\AA} using only the $\boldsymbol{\Gamma}$-point. The long-ranged electrostatic interactions were decoupled from the periodic images for the molecules.\cite{Bloechl1995} The cell size was converged so that no artificial interactions between periodic images are taken into account for molecular systems. Wave function and geometry optimizations were performed using the Car-Parrinello ab-initio molecular dynamics method\cite{Car1985} with a friction term which quenches the system to the ground state. This enabled efficient optimizations of the atomic positions in the unit cell. The computational costs of geometry optimizations increased only by roughly a factor of 2 compared to single-point calculations depending on the initial structure. For metallic systems, the Mermin functional\cite{Mermin1965} was applied to treat variable occupations of the one-electron energy eigenstates. The total energy was minimized up to a numerical convergence of $10^{-5}\,E_\mathrm{H}$ for the given settings.
PBE$_\mathrm{PAW}$ calculations were performed with the CP-PAW code as well using the same settings with the exception that all HF mixing factors were set to zero.
\\ 
The PBE, HSE06, and PBE0 calculations were performed using the Fritz-Haber-Institute ab initio molecular simulations (FHI-aims) package (version 160328{\_}3)\cite{Blum2009} which is an all-electron electronic structure code with numeric atom-centered basis functions. Again, a collinear treatment of the spin-polarization was applied. The default light basis set of FHI-aims was used which achieves a finite basis set error of less than 0.04\,eV per atom for energy differences like formation energies. A $\boldsymbol{\Gamma}$-centered $\mathbf{k}$-point grid was used for periodic systems. The density of the $\mathbf{k}$-point grid was the same as in the CP-PAW calculations. The error of the finite $\mathbf{k}$-point grid is less than 0.001\,eV per atom for energy differences unless metallic systems are present, then the error can be up to 0.04\,eV per atom. Molecular systems were calculated in a non-periodic environment which saved computational resources while full numerical consistency of the settings with periodic calculations was maintained. Geometry optimizations were performed using the Broyden{-}Fletcher{-}Goldfarb{-}Shanno algorithm\cite{Broyden1970, Fletcher1970, Goldfarb1970, Shanno1970} up to a numerical convergence of 0.001\,eV of the system's total energy whereby the used forces provide an accuracy of $2\cdot10^{-4}$\,eV/{\AA}. An exception were the $\upalpha$-Mn calculations where the total energy and forces were only converged up to 0.01\,eV and $2\cdot10^{-3}$\,eV/{\AA}, respectively. The total energies themselves were converged in every iteration of the geometry optimizations in a self-consistent field procedure to a numerical accuracy of $10^{-5}\,\mathrm{eV}$. For metallic systems, the zero-broadening corrected energies were used. Further details of the FHI-aims calculations are given in the SI.

\section{Results and Discussion}

\subsection{Magnetic Order}\label{sec_magnetic_order}

\begin{figure}[b!]
\centering
\includegraphics[width=\columnwidth]{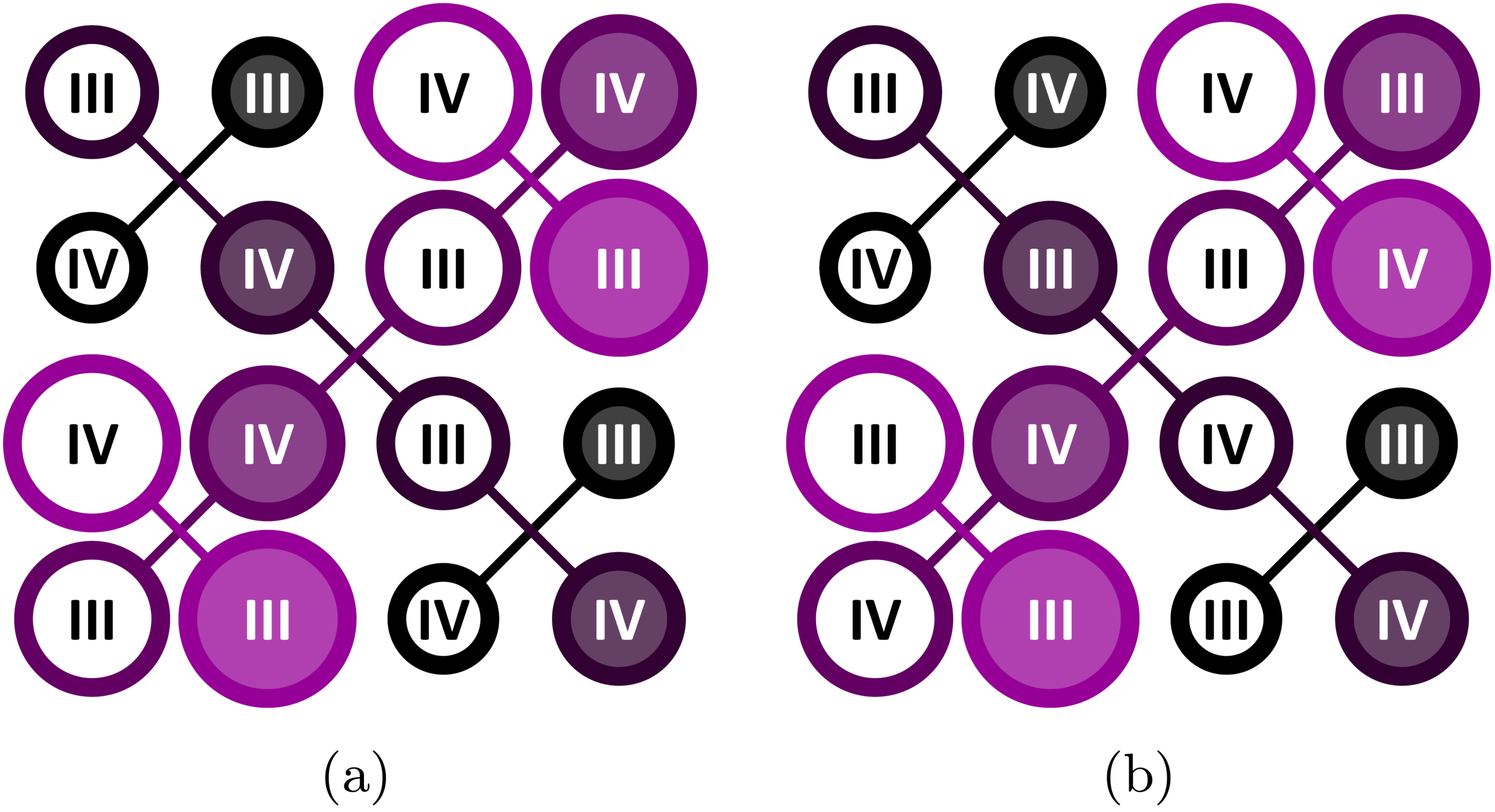}
\caption{The lowest-energy oxidation and spin states of the Mn atoms in LiMn$_2$O$_4$ obtained by PBE0r (a) as well as by PBE0 and HSE06 (b). The size and color represent the position of Mn in the $z$ direction of the unit cell: from large/bright (top layer) to small/dark (bottom layer). The Roman numerals correspond to the oxidation states. Filled and empty circles represent the two spin channels. The lines define the $\left(\text{MnO}_6\right)_4$ tetrahedra.}\label{fig_limn2o4_charge_spin}
\end{figure}

The experimentally determined atomic structures referenced in Sec.\ \ref{sec_introduction} were taken as starting geometry of the electronic structure calculations. The structures were optimized by the individual DFT functionals under the constraint of fixed lattice vectors which were taken from experimental data. For most of the systems the initial spin configurations were taken from the references mentioned in Sec.\ \ref{sec_introduction} or they were derived by projecting the given non-collinear spins onto a collinear arrangement. Otherwise, a search for the minimum energy spin configuration was performed. The procedure for identifying the minimum energy spin configuration as well as the classification of the hs-Mn$^\text{III}$ and Mn$^\text{IV}$ ions is described in the SI. The initial spins were fully optimized including a possible reordering -- but no spin-flips -- in the subsequent electronic structure calculations.
\\
The spin configuration of cubic LiMn$_2$O$_4$ was investigated in several previous theoretical and experimental studies\cite{Ouyang2009, Sugiyama1997, Rodriguez-Carvajal1998, Tomeno2001, Jang2000} which showed that the corner-sharing $\left(\text{MnO}_6\right)_4$ tetrahedra, formed for example by the four upper left Mn ions shown in FIG.\ \ref{fig_limn2o4_charge_spin}, generally contain two Mn$^\text{III}$ and two Mn$^\text{IV}$ ions. Furthermore, an antiferromagnetic long-range order at low temperatures was proposed. Our calculations confirm this result in that a ferromagnetic structure is energetically less stable then a configuration with an overall zero magnetic moment. 
\\
FIG.\ \ref{fig_limn2o4_charge_spin} illustrates the lowest-energy oxidation and spin states of the Mn ions calculated by the hybrid functionals. First of all, we note that there are many other configurations with very similar energies differing only by a few meV per atom. While the PBE0 and HSE06 calculations yield the same configuration, the result obtained in the PBE0r calculations differs slightly. Both configurations have in common that the spins of the Mn ions are ordered in antiferromagnetically coupled (100) planes (planes of filled and empty circles in FIG.\ \ref{fig_limn2o4_charge_spin}). Moreover, in each plane there is an equal number of Mn$^\text{III}$ and Mn$^\text{IV}$ ions. Additionally, all $\left(\text{MnO}_6\right)_4$ tetrahedra consist of two Mn$^\text{III}$ and two Mn$^\text{IV}$ ions. This is in agreement with the previous studies mentioned above. However, the distribution of Mn atoms with different oxidation states within the network of tetrahedra is not the same in PBE0 and HSE06 on the one hand and PBE0r on the other hand. Still, the second lowest minimum found by HSE06 is the minimum of PBE0r. The total energy difference between these two spin configurations is only 0.002\,eV per atom, which is within the remaining uncertainty of the hybrid functionals. The PBE0r minimum configuration is also among the energetically lowest configurations of PBE0. The PBE0 energy difference between the PBE0r minimum configuration and the PBE0 minimum is 0.003\,eV per atom and thus very small. In conclusion, the energy differences are one order of magnitude smaller than the error of the finite basis set (0.01\,eV per atom) in all of these calculations. The latter is in the range of the experimental uncertainty for formation enthalpies.\cite{Wang2005, Cupid2015} Therefore, the two magnetic orders in FIG.\ \ref{fig_limn2o4_charge_spin} can be considered as degenerate within the given accuracy.
\\
Because both spin configurations show antiferromagnetically coupled (100) planes with an equal number of Mn$^\text{III}$ and Mn$^\text{IV}$ ions and all $\left(\text{MnO}_6\right)_4$ tetrahedra also contain two Mn$^\text{III}$ and two Mn$^\text{IV}$ ions, there is no fundamental difference in the description of the magnetic order by PBE0, HSE06, and PBE0r. Therefore, and because the energies of all configurations are very similar, in the remaining part of this work the PBE0r minimum configuration is used for all calculations.
\\
\begin{figure}[b!]
\centering
\includegraphics[width=\columnwidth]{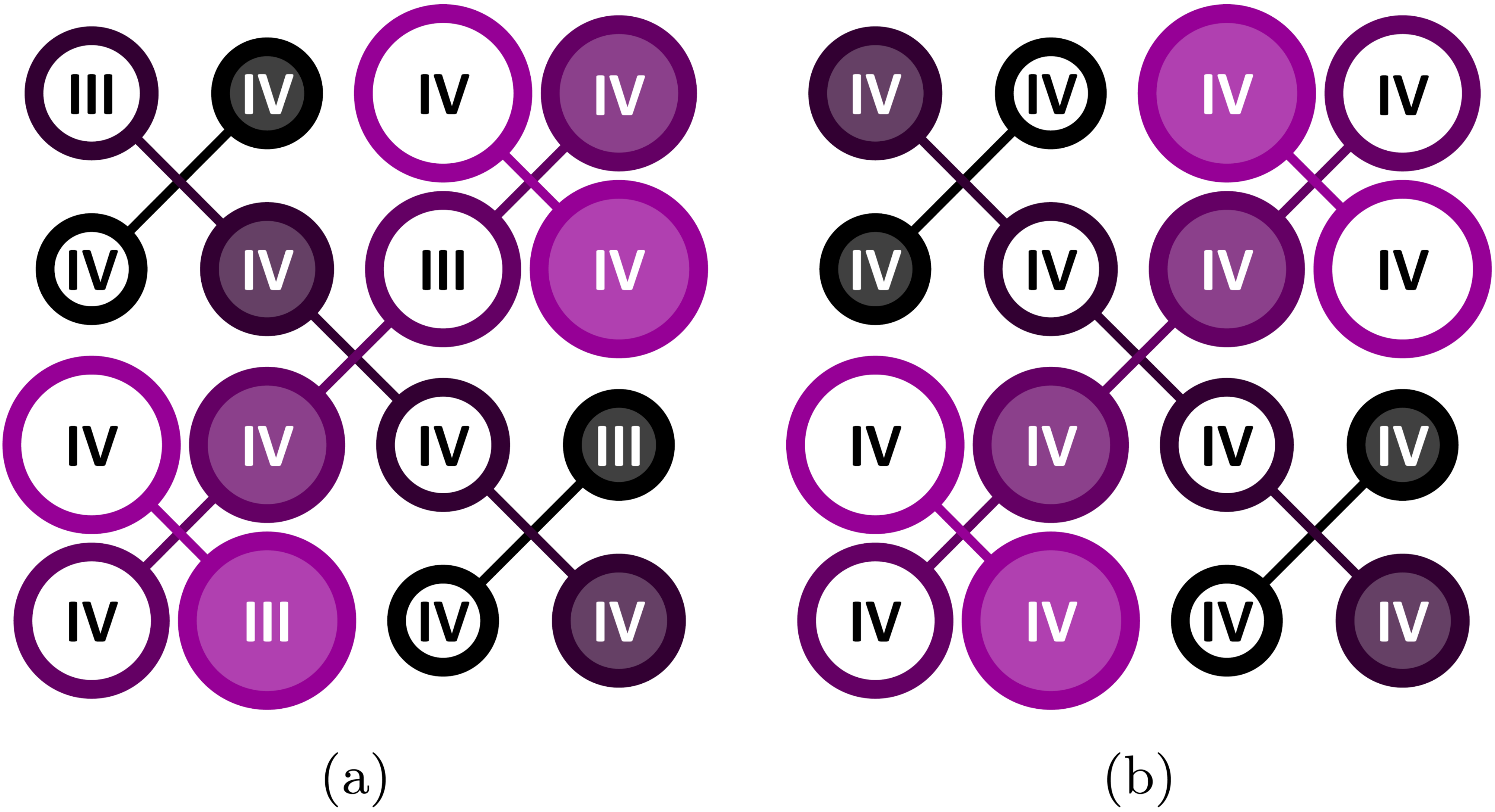}
\caption{The lowest-energy oxidation and spin states of the Mn atoms calculated consistently by all hybrid functionals for (a): $\uplambda$-Mn$_2$O$_4$ and (b): Li$_{0.5}$Mn$_2$O$_4$. The size and color represent the position of Mn in the $z$ direction of the unit cell: from large/bright (top layer) to small/dark (bottom layer). The Roman numerals correspond to the oxidation states. Filled and empty circles represent the two spin channels. The lines define the $\left(\text{MnO}_6\right)_4$ tetrahedra.}\label{fig_mn2o4_li05mn2o4_charge_spin}
\end{figure}
Next, the oxidation states and spin directions in the minimum energy configurations of Li$_{0.5}$Mn$_2$O$_4$ and $\uplambda$-Mn$_2$O$_4$ are investigated. For both systems, the HSE06, PBE0, and PBE0r fully agree with each other (FIG.\ \ref{fig_mn2o4_li05mn2o4_charge_spin}). The spin directions of the Mn ions in Li$_{0.5}$Mn$_2$O$_4$ are oriented in the same way as in LiMn$_2$O$_4$. As a consequence of the reduced Li content the number of Mn$^\text{III}$ ions in Li$_{0.5}$Mn$_2$O$_4$ is reduced to four per unit cell resulting in the ratio Mn$^\text{III}$:Mn$^\text{IV}$ of 1:3 in each spin plane. The energetically lowest order of the Mn spins is different in the case of $\uplambda$-Mn$_2$O$_4$ compared to those of the lithiated compounds. The planes of equal spin are replaced by an alternating sequence in which the spin direction changes every second Mn ion. As in the case of LiMn$_2$O$_4$ also for Li$_{0.5}$Mn$_2$O$_4$ and $\uplambda$-Mn$_2$O$_4$ several magnetic orders with energies differing only in the order of a few meV per atom exist. The resulting minimum energy spin configurations of the PBE0r geometry optimized structures are given in the SI for all benchmark systems considered in this study.

\subsection{Formation Energies}\label{sec_formation}

Arguably, the most important benchmark property is the relative stability of different structures. In particular, formation energies $\Delta E_\mathrm{f}$ can be compared to other levels of theory and to experimental formation enthalpies $\Delta H_\mathrm{f}$ to judge the quality of a given exchange-correlation functional. The formation energy is defined as the difference between the total energy of a given structure and the sum of the total energies of the elements in their reference modification, i.e.\  H$_{2}\mathrm{(g)}$, O$_{2}\mathrm{(g)}$, bulk bcc Li, and bulk $\upalpha$-Mn in the present case. In contrast to formation energies, experimental formation enthalpies also include the zero point energy. Additionally, formation enthalpies are determined at standard conditions, i.e.\ at a temperature of 298.15\,K and a pressure of 1\,bar. However, the additional contributions are typically small compared to potential energy differences of different chemical compounds or structures. Therefore, comparing formation energies $\Delta E_\mathrm{f}$ and standard formation enthalpies $\Delta H_\mathrm{f}^\circ$ is a reasonable and frequently used approximation, which we will also employ here. 
\\
\begin{table*}[tb]
\centering
\caption{Calculated formation energies $\Delta E_\mathrm{f}$ and experimental standard formation enthalpies $\Delta H_\mathrm{f}^\circ$ in eV per formula unit for the investigated benchmark systems. The structures were optimized under the constraint of fixed lattice vectors which were taken from experimental data. The mean absolute error ($\mathrm{MAE}=\tfrac{1}{n_\mathrm{exp}}\sum_{i=1}^{n_\mathrm{exp}}\left|{\Delta E_\mathrm{f}}_i-{\Delta H_\mathrm{f}^\circ}_i\right|$) is calculated using the $n_\mathrm{exp}$ systems for which experimental data are given excluding H$_2$O(g). The mean absolute error per atom ($\mathrm{MAE\,/\,atom}=\tfrac{1}{n_\mathrm{exp}}\sum_{i=1}^{n_\mathrm{exp}}\tfrac{1}{{n_\mathrm{atoms}}_i}\left|{\Delta E_\mathrm{f}}_i-{\Delta H_\mathrm{f}^\circ}_i\right|$) uses for each system the error per atom instead of the error per formula unit which includes $n_\mathrm{atoms}$ atoms.}
\begin{tabular}{lrrrrrr}
\hline
 &\multicolumn{5}{c}{$\Delta E_\mathrm{f}$} & $\Delta H_\mathrm{f}^\circ$ \\
System & PBE & PBE$_\mathrm{PAW}$ & PBE0 & HSE06 & PBE0r &  Exp.\\
\hline
H$_2$O(g) & $-$2.49 & $-$2.52 & $-$2.64 & $-$2.63 & $-$2.38 & $-$2.51\cite{Robie1979} \\
Li$_2$O$_2$ & $-$5.79 & $-$6.18 & $-$5.99 & $-$5.98 & $-$5.91 & $-$6.56\cite{JANAF1963} \\
Li$_2$O & $-$5.60 & $-$6.00 & $-$5.88 & $-$5.87 & $-$5.78 & $-$6.21\cite{Robie1979} \\
MnO & $-$2.54 & $-$2.58 & $-$4.49 & $-$4.46 & $-$3.75 & $-$3.99\cite{Robie1979} \\
Mn$_3$O$_4$ & $-$11.56 & $-$11.61 & $-$16.00 & $-$16.05 & $-$14.30 & $-$14.38\cite{Robie1979} \\
$\upalpha$-Mn$_2$O$_3$ & $-$8.35 & $-$8.35 & $-$10.91 & $-$10.98 & $-$9.95 & $-$9.94\cite{Robie1979} \\
$\upbeta$-MnO$_2$ & $-$5.07 & $-$5.07 & $-$5.51 & $-$5.61 & $-$5.51 & $-$5.39\cite{Robie1979} \\
LiMnO$_2$ & $-$7.48 & $-$7.56 & $-$8.98 & $-$8.99 & $-$8.31 & $-$8.70\cite{Wang2005} \\
$\uplambda$-Mn$_2$O$_4$ & $-$9.81 & $-$9.81 & $-$10.60 & $-$10.71 & $-$10.76 & --- \\
Li$_{0.5}$Mn$_2$O$_4$ & $-$11.58 & $-$11.65 & $-$12.81 & $-$12.91 & $-$12.55 & --- \\
LiMn$_2$O$_4$ & $-$13.15 & $-$13.26 & $-$14.91 & $-$14.98 & $-$14.30 & $-$14.32\cite{Cupid2015} \\
Li$_2$Mn$_2$O$_4$ & $-$15.18 & $-$15.36 & $-$18.07 & $-$18.09 & $-$16.78 & $-$17.34\cite{Wang2005} \\
\hline
MAE & 1.35 & 1.21 & 0.63 & 0.67 & 0.28 & \\
MAE\,/\,atom & 0.31 & 0.27 & 0.15 & 0.16 & 0.08 & \\
\hline
\end{tabular}
\label{tab_formation_energy_1}
\end{table*}
\begin{figure*}[tb]
\centering
\includegraphics[width=\textwidth]{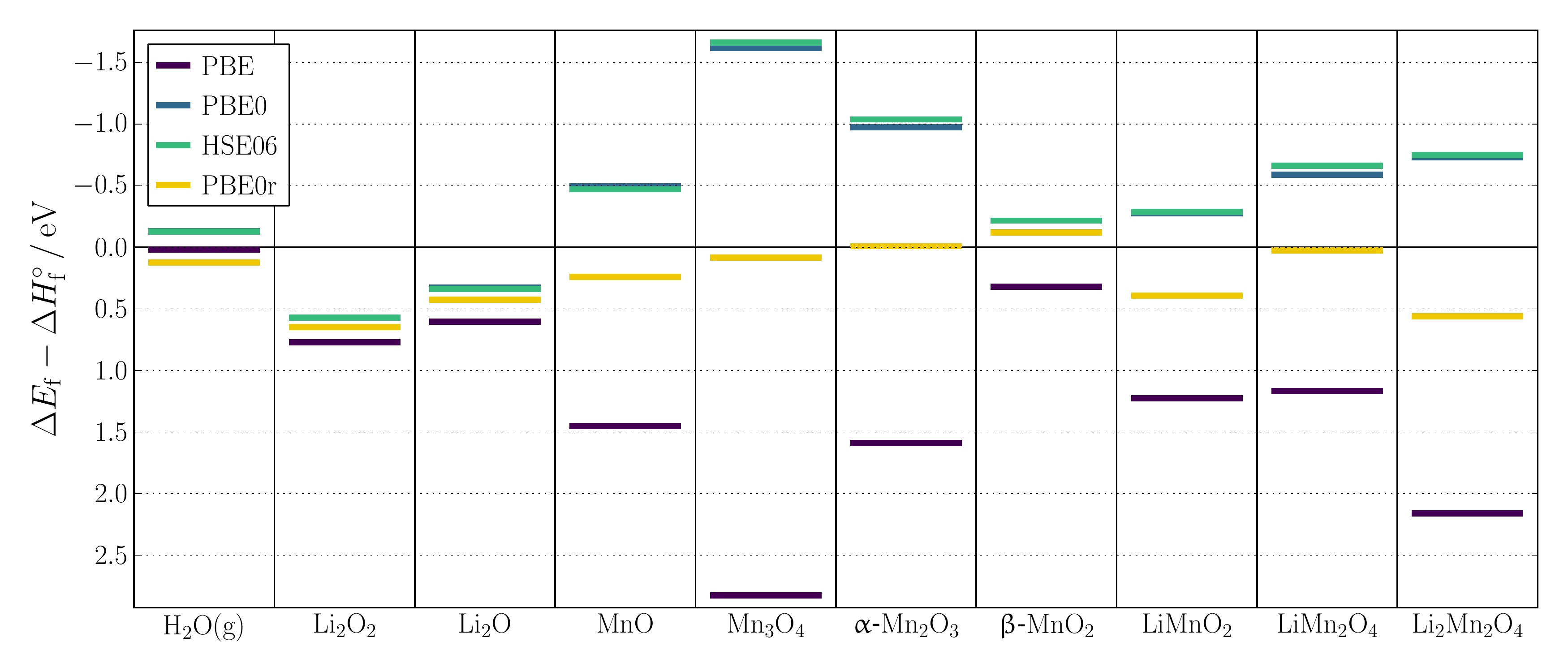}
\caption{Differences between calculated formation energies and experimental standard formation enthalpies $\Delta E_\mathrm{f}-\Delta H_\mathrm{f}^\circ$ in eV per formula unit for the benchmark systems obtained by the investigated functionals. The energy axis is inverted to show overestimated formation energies above the zero line and underestimated formation energies below. All calculated formation energy values and standard formation enthalpies are given in TABLE \ref{tab_formation_energy_1}.}\label{fig_E}
\end{figure*}
The calculated formation energies and the corresponding experimental formation enthalpies of the benchmark systems are given in TABLE \ref{tab_formation_energy_1}. The internal structures of the benchmark systems were optimized by the respective functionals under the constraint of fixed lattice vectors. The unit cell parameters were taken from the experimental data cited in Sec.\ \ref{sec_introduction}. The deviations between the theoretical and experimental results are plotted in FIG.\ \ref{fig_E}. A comparison to theoretical results from previous studies is given in the SI.
\\
First, the agreement between FHI-aims and CP-PAW calculations is investigated by performing PBE calculations with both codes. As shown in the columns PBE (FHI-aims) and PBE$_\mathrm{PAW}$ (CP-PAW) of TABLE \ref{tab_formation_energy_1}, the agreement for the formation energies of the manganese oxides and water is very good. However, the formation energies of the lithium oxides show larger deviations of up to 0.13\,eV per atom. This is mainly related to the description of the Li atoms, which is also visible in the deviations between the PBE formation energies for the lithium manganese oxides calculated with both codes. Also in this case the discrepancies increase with larger Li contents. Tests have shown that if the 1s electrons of Li would also have been treated as valence electrons in the CP-PAW calculations, the deviation, for example, for the Li$_2$O formation energy could have be reduced by 0.05\,eV per atom. The PBE value from CP-PAW for the formation energy of Li$_2$O$_2$, which deviates by 0.10\,eV per atom from the value obtained by FHI-aims, agrees within 0.02\,eV per atom with the Vienna Ab initio Simulation Package (VASP).\cite{Chan2011, Kresse1996} For all benchmark system except the lithium oxides the maximum deviation of the formation energies is 0.02\,eV per atom, which is within the basis set error.
\\
Irrespective of the DFT code, the PBE formation energies of all (lithium) manganese oxides are much smaller, i.e.\ less negative, than the experimental enthalpies. These systematic deviations have previously been explained by the self-interaction error of GGA functionals.\cite{Cramer2009} (Lithium) manganese oxides are highly correlated systems as the Mn d electrons are strongly localized. The size of the Mn 3d valence orbitals is similar to the size of the 3s and 3p core orbitals which leads to a strong Coulomb interaction. Those systems are typically most severely affected by the self-interaction error. Specifically, the energy of spatially localized states is unphysically increased by the interaction of an electron with its own charge density. Therefore, delocalized charge distributions, i.e.\ metallic behavior, are favored. As a consequence, metallic Mn is more favored than oxidic Mn and consequently the formation energies of the (lithium) manganese oxides are smaller. The introduction of exact exchange contributions aims to reduce the self-exchange error, which results in a more accurate description of localized states. In conclusion, the formation energies of the (lithium) manganese oxides should be larger in case of hybrid functionals compared to the PBE GGA functional, which is consistent with our improved results for the oxide benchmark systems reported in TABLE \ref{tab_formation_energy_1}.
\\
The results of the PBE0 and HSE06 functionals are very similar. Therefore, the exclusion of the long-range exact exchange contributions seems to be a good approximation leading only to small errors. However, both functionals overestimate the formation energies of all (lithium) manganese oxides compared to the experimental values, while PBE underestimates the formation energies. Therefore, a HF mixing factor between 0 and 25{\%} should improve the agreement with experiment. This is the case for PBE0r, which uses only 9{\%} HF on-site exchange for Mn while off-site exchange terms are described by GGA only, and indeed we find a better agreement with the experimental data of the (lithium) manganese oxides as shown in FIG.\ \ref{fig_E}. The freedom of choosing the amount of on-site exact exchange allows -- to a certain extent -- to compensate for errors introduced by the local approximation in PBE0r. The formation energies of the water molecule and the lithium oxides are not as much affected by the inclusion of HF exchange. The PBE0r formation energies for the oxides are systematically higher than the PBE results but smaller than the PBE0 data. This is in accordance with the intermediate HF mixing factor. The formation energy of water is an exception of this trend. The water molecule is covalently bonded. Here, the restriction to on-site exchange terms seems to miss relevant contributions.
\\
Compared to experiment, the PBE0r formation energy of the H$_2$O monomer is underestimated by 0.13\,eV, while it is overestimated in case of PBE0 and HSE06 by about 0.13 and 0.12\,eV, respectively. The experimental enthalpy is 0.03\,eV smaller at 0\,K compared to the standard value at 298\,K\cite{Ruscic2006} given in TABLE \ref{tab_formation_energy_1}. 
However, the experimental formation enthalpy includes a reduction by the zero point energy\cite{Feller2014} while the calculated formation energies in TABLE \ref{tab_formation_energy_1} have not been corrected for the zero point energy and are thus expected to be too high. Since the contribution of the experimental zero point energy is about 0.24\,eV\cite{Webbook_H2, Webbook_O2, Webbook_H2O}, the experimental formation energy at 0\,K without zero point energy correction is $-$2.72\,eV. Thus, the PBE0 and HSE06 results are more accurate for the formation energy of the H$_2$O monomer. The less accurate result of PBE0r could be caused by the lack of off-site exact exchange, which is important for the covalent bonds in H$_2$, O$_2$, and H$_2$O. The zero point energies are typically much smaller if no H atoms are present in the system. 
\begin{table}[H]
\centering
\caption{Formation energies $\Delta E_\mathrm{ox}$ and experimental standard formation enthalpies $\Delta H_\mathrm{ox}^\circ$ for the investigated benchmark systems calculated with respect to the oxides Li$_2$O, MnO, and $\upbeta$-MnO$_2$ in eV per formula unit. The structures were optimized under the constraint of fixed lattice vectors, which were taken from experimental data. The calculation of the MAE is performed as described in FIG.\ \ref{fig_E}.}
\begin{tabular}{lrrrr}
\hline
 &\multicolumn{3}{c}{$\Delta E_\mathrm{ox}$} & $\Delta H_\mathrm{ox}^\circ$ \\
System & PBE0 & HSE06 & PBE0r &  Exp.\\
\hline
Mn$_3$O$_4$ & $-$1.51 & $-$1.51 & $-$1.29 & $-$1.01\cite{Robie1979} \\
$\upalpha$-Mn$_2$O$_3$ & $-$0.91 & $-$0.91 & $-$0.69 & $-$0.56\cite{Robie1979} \\
LiMnO$_2$ & $-$1.04 & $-$1.02 & $-$0.79 & $-$0.91\cite{Robie1979,Wang2005} \\
LiMn$_2$O$_4$ & $-$1.45 & $-$1.41 & $-$1.27 & $-$1.14\cite{Robie1979,Cupid2015} \\
Li$_2$Mn$_2$O$_4$ & $-$2.19 & $-$2.16 & $-$1.74 & $-$1.75\cite{Robie1979,Wang2005} \\
\hline
MAE & 0.35 & 0.33 & 0.13 & \\
MAE\,/\,atom & 0.05 & 0.05 & 0.02 & \\
\hline
\end{tabular}
\label{tab_formation_energy_2}
\end{table}
The PBE0r approximation of on-site exchange should work best for oxides. To check this hypothesis, formation energies $\Delta E_\mathrm{ox}$ and experimental standard formation enthalpies $\Delta H_\mathrm{ox}^\circ$ have also been calculated with respect to a reference of the oxides Li$_2$O, MnO, and $\upbeta$-MnO$_2$ instead of the elemental reference states for the investigated benchmark systems (TABLE \ref{tab_formation_energy_2}). Therefore, these energy difference do not include information from the covalently bonded H$_2$, O$_2$, and H$_2$O and metallic Li and $\upalpha$-Mn, which are most critical in case of the PBE0r functional. Again, the results of PBE0 and HSE06 are overall very similar with a slightly smaller mean absolute error (MAE) of the HSE06 results compared to experimental data. As expected, the PBE0r results now show a much reduced MAE of 0.13\,eV which is about 2.5 times smaller than the MAE for the other two hybrid functionals. In general, the MAEs for the formation energies with respect to the oxides are almost half of the MAEs for the formation energies from the elements (TABLE \ref{tab_formation_energy_1}) in the case of all hybrid functionals.
\\
In summary, the MAE of the formation energies from the elements for all benchmark systems can be reduced from 1.35\,eV (PBE) to 0.63\,eV (PBE0) by the inclusion of 25{\%} HF exchange confirming that the inclusion of exact exchange is very important for these systems. However, this increases the average computation time of the benchmark systems by a factor of about 25. The MAE is only very little affected by the neglect of the long-range exact exchange terms (PBE0 vs.\ HSE06). This decreases the computational effort of HSE06 compared to PBE0 by a factor of 0.8. The MAE can further be decreased to 0.28\,eV (PBE0r) for these benchmark systems if smaller HF mixing factors are used. For the formation energy with respect to reference oxides the MAE of the PBE0r functional is only 0.13\,eV providing the best agreement with the experimental data and demonstrarting a very good description of energy differences. However, the main advantage of the PBE0r functional is that the average computation time per iteration of the benchmark systems compared to PBE$_\mathrm{PAW}$ is only increased by a factor of 1.3 while the error is reduced by a factor of 4.

\subsection{Intercalation Potential}

An important property of the Li$_x$Mn$_2$O$_4$ spinel is the possibility to remove or incorporate Li atoms by (de)intercalation. This enables the application as cathode material for lithium ion batteries. Experiments show that the standard electrochemical potential of a Li/Li$_x$Mn$_2$O$_4$ cell is about 4\,V vs.\ Li/Li$^+$ in the range of $0\leq x\leq1$ and 3\,V in the range of $1<x\leq2$.\cite{Gummow1994}
\\
The intercalation potential $\Delta G_\mathrm{p}$ can be calculated by
\begin{align}
\Delta G_\mathrm{p}=\dfrac{(x_2-x_1)G_\mathrm{Li}+G_{\mathrm{Li}_{x_1}\mathrm{Mn}_2\mathrm{O}_4}-G_{\mathrm{Li}_{x_2}\mathrm{Mn}_2\mathrm{O}_4}}{(x_2-x_1)}\ ,
\end{align}
where $x_1$ and $x_2$ specify the Li content of the Li$_x$Mn$_2$O$_4$ structure.\cite{Aydinol1997} It corresponds to the Gibbs free energy difference of the reaction
\begin{align}
\mathrm{Li}_{x_2}\mathrm{Mn}_2\mathrm{O}_4 \rightleftharpoons \mathrm{Li}_{x_1}\mathrm{Mn}_2\mathrm{O}_4+(x_2-x_1)\mathrm{Li}
\end{align}
divided by $(x_2-x_1)$. The Gibbs free energy $G$ can be approximated by the total energy $E$ when neglecting the influence of zero point energies and entropic contributions at finite temperatures. The resulting error of this approximate treatment is typically small for this reaction since the change in the neglected terms is much smaller than the change in the total energy $E$ during this reaction.\cite{Aydinol1997} If the intercalation potential of Li$_2$Mn$_2$O$_4$/LiMn$_2$O$_4$ is calculated from the experimental formation enthalpies given in TABLE \ref{tab_formation_energy_1}, the result is an intercalation potential of 3.02\,eV which matches the electrochemically determined data given in TABLE \ref{tab_V} very well.
\\
\begin{table}[b!]
\centering
\caption{Calculated intercalation potentials $\Delta E_\mathrm{p}$ and experimentally measured standard electrochemical potentials vs. Li/Li$^+$ $E_\mathrm{p}^0$ in eV.}
\begin{tabular}{lrrrrr}
\hline
Potential & PBE & PBE0 & HSE06 & PBE0r & Exp. \\
\hline
Li$_{0.5}$Mn$_2$O$_4$/$\uplambda$-Mn$_2$O$_4$ & 3.54 & 4.43 & 4.39 & 3.59 & 4.1\cite{Xia1996} \\
LiMn$_2$O$_4$/Li$_{0.5}$Mn$_2$O$_4$ & 3.14 & 4.19 & 4.15 & 3.49 & 4.0\cite{Xia1996} \\
Li$_2$Mn$_2$O$_4$/LiMn$_2$O$_4$ & 2.03 & 3.16 & 3.11 & 2.48 & 3.0\cite{Peramunage1998} \\
\hline
MAE & 0.80 & 0.23 & 0.18 & 0.51 & \\
MAE\,/\,atom & 0.11 & 0.03 & 0.03 & 0.07 & \\
\hline
\end{tabular}
\label{tab_V}
\end{table}
The experimentally measured standard electrochemical potentials vs. Li/Li$^+$ $E_\mathrm{p}^0$ are in between the calculated PBE and PBE0 intercalation potentials $\Delta E_\mathrm{p}$ (TABLE \ref{tab_V}). The MAE of the three PBE0 intercalation potentials with respect to to the experiment is 0.23\,eV and thus much smaller than the MAE of PBE, which is 0.80\,eV. The HSE06 results exhibiting an MAE of 0.18\,eV are a little closer to the experimental data than PBE0. As expected, the PBE0r values (MAE 0.51\,eV) are in between the  PBE and PBE0 data, but only slightly better than PBE and clearly less accurate than PBE0 and HSE06. Still, the trend of the intercalation potential as a function of the Li content $x$ is described reliably by PBE0r in contrast to PBE which predicts the relative differences among the potentials at different lithium contents inaccurately. PBE0 and HSE06 describe the relative differences well but not as good as PBE0r whose results show a relatively constant underestimation of approximately 0.5\,eV for all potentials. 

\subsection{Structural Properties}\label{sec_structure}

Up to this point, all properties have been obtained from optimized atomic positions using the experimental lattice parameters. A natural next question is how well the tested functionals are able to describe the structural and structure-related properties such as the equilibrium lattice constants and bulk moduli. To answer this question, we used the Birch-Murnaghan equation of state~\cite{Birch1947, Murnaghan1937}
\begin{align}
\begin{split}
U={}&U_0+\tfrac{9}{16}B_0V_0\left[\left(\tfrac{V_0}{V}\right)^{\tfrac{2}{3}}-1\right]^2\\
&\left\{B_0^\prime\left[\left(\tfrac{V_0}{V}\right)^{\tfrac{2}{3}}-1\right]-4\left(\tfrac{V_0}{V}\right)^{\tfrac{2}{3}}+6\right\},
\end{split}
\end{align}
to derive the equilibrium energy $U_0$, equilibrium volume $V_0$, the bulk modulus at zero pressure $B_0$, and its pressure derivative $B_0^\prime$. To reduce the search space, in case of non-cubic cells we have kept the ratios of the lattice parameters and scaled the experimental lattice constants isotropically by factors in the range from 0.95 to 1.05 in steps of 0.01. Again, for each calculation the atomic positions in the cells were optimized under the constraint of the respective lattice constants. 
\\
Approximating the internal energy $U$ by the total energy $E$, the Birch-Murnaghan equation of state describes the $E(V)$ relation for all periodic benchmark systems very well.
The formation energies using the oxides Li$_2$O, MnO, and $\upbeta$-MnO$_2$ as reference were calculated using the resulting equilibrium energies $E_0$ from the Birch-Murnaghan equation of state for each hybrid functional. The values are given in the SI. These energies deviate only very little from the DFT data using the experimental lattice constants, which are given in TABLE \ref{tab_formation_energy_2}. The MAE of the energies per formula unit changes only by 0.009\,eV for PBE0, $-$0.001\,eV for HSE06, and $-$0.013\,eV for PBE0r.
\\
\begin{table}[t!]
\centering
\caption{Relative deviations between the experimental equilibrium lattice constants and the calculated values using the examined functionals in {\%}.}
\begin{tabular}{lrrrr}
\hline
System & PBE & PBE0 & HSE06 & PBE0r \\
\hline
Li & $-$0.9 & 0.0 & $-$0.3 & $-$2.1 \\
Li$_2$O$_2$ & 0.9 & $-$0.4 & $-$0.4 & $-$0.1 \\
Li$_2$O & 0.8 & $-$0.5 & $-$0.4 & $-$0.5 \\
MnO & 0.1 & $-$0.1 & 0.1 & 1.0 \\
Mn$_3$O$_4$ & 0.3 & 0.0 & 0.1 & 1.3 \\
$\upalpha$-Mn$_2$O$_3$ & 0.5 & 0.0 & 0.1 & 1.5 \\
$\upbeta$-MnO$_2$ & 0.3 & $-$0.7 & $-$0.6 & 0.7 \\
LiMnO$_2$ & 1.1 & 0.1 & 0.2 & 1.1 \\
$\uplambda$-Mn$_2$O$_4$ & 0.7 & $-$0.4 & $-$0.2 & 1.2 \\
Li$_{0.5}$Mn$_2$O$_4$ & 0.2 & $-$0.1 & 0.0 & 1.1 \\
LiMn$_2$O$_4$ & $-$0.2 & $-$0.1 & 0.1 & 0.8 \\
Li$_2$Mn$_2$O$_4$ & 0.7 & 0.0 & 0.1 & 0.9 \\
\hline
Mean Error & 0.6 & 0.2 & 0.2 & 1.0 \\
\hline
\end{tabular}
\label{tab_lattice_constant}
\end{table}
TABLE \ref{tab_lattice_constant} shows the relative deviations between the calculated equilibrium lattice constants and the experimental data. In contrast to the formation energies, the PBE equilibrium lattice constants do not show large deviations from the experimental data with an overall relative error of only 0.6{\%}, which is in agreement with the general finding that structural properties are well described already at the GGA level. PBE0 and HSE06, which have a tendency to only very slightly underestimate the lattice parameters, yield an even smaller error as low as 0.2{\%} with respect to experiment. This is reasonable as the experimental results were determined at finite temperature, and should be somewhat higher than the theoretical results obtained at 0\,K due to thermal expansion. 
\\
PBE0r underestimates in particular the lattice constants of Li by 2.1{\%}. The lattice constants of the lithium oxides are described with an accuracy comparable to the other hybrid functionals. The equilibrium lattice constants of all given (lithium) manganese oxides are overestimated between 0.7 and 1.5{\%} thus covering a range of 0.8{\%}.  PBE0 and HSE06 span exactly the same range of 0.8{\%} (from $-$0.7 to 0.1 and from $-$0.6 to 0.2, respectively). Thus, the relative differences are described similarly by all hybrid functionals. Still, there is a constant shift of about 1{\%} to larger lattice parameters with PBE0r compared to PBE0 and HSE06. Consequently, this leads to a mean relative error of approximately 1.0{\%} for PBE0r which is still a very good agreement. We attribute the overestimation of the lattice constants to the neglect of off-site HF exchange terms which we expect to strengthen bonds. The equilibrium bond lengths for the molecular systems are reported in the SI.
\\
The results for $\upalpha$-Mn are not listed in TABLE \ref{tab_lattice_constant} and \ref{tab_bulk_modulus} because no minimum is observed in the isotropic compression/expansion range from 0.95 to 1.05 using any of the hybrid functionals. If the equilibrium lattice constant is calculated by PBE with collinear spin, the outcome is an underestimation of 2.7{\%}, which is very similar to a previous PBE study employing non-collinear spin.\cite{Hobbs2003} The error of PBE0, HSE06, and PBE0r could originate from the restriction to a collinear spin arrangement in our calculations, as $\upalpha$-Mn has a complex non-collinear magnetic electronic structure, whose characterization is beyond the scope of the present work due to the large computational effort. 
\begin{table}[H]
\centering
\caption{Calculated and experimental bulk moduli $B_0$ in GPa for the benchmark systems. The bulk modulus of $\upbeta$-MnO$_2$ has not been included in the calculation of the MAE with respect to experiment because of the large uncertainty in the experimental value.}
\begin{tabular}{lrrrrr}
\hline
System & PBE & PBE0 & HSE06 & PBE0r & Exp. \\
\hline
Li & 13 & 13 & 13 & 14 & 12\cite{Trivisonno1961} \\
Li$_2$O$_2$ & 72 & 79 & 79 & 76 & --- \\
Li$_2$O & 74 & 82 & 81 & 80 & 82\cite{Hull1988} \\
MnO & 124 & 162 & 161 & 148 & 154\cite{Oliver1969} \\
Mn$_3$O$_4$ & 127 & 149 & 149 & 131 & 133\cite{Darul2013} \\
$\upalpha$-Mn$_2$O$_3$ & 140 & 173 & 173 & 151 & 169\cite{Yamanaka2005} \\
$\upbeta$-MnO$_2$ & 224 & 263 & 262 & 229 & $260-280$\cite{Haines1995} \\
LiMnO$_2$ & 111 & 127 & 127 & 112 & --- \\
$\uplambda$-Mn$_2$O$_4$ & 100 & 119 & 118 & 100 & --- \\
Li$_{0.5}$Mn$_2$O$_4$ & 108 & 126 & 126 & 108 & --- \\
LiMn$_2$O$_4$ & 118 & 129 & 133 & 126 & 119\cite{Lin2011} \\
Li$_2$Mn$_2$O$_4$ & 114 & 129 & 129 & 115 & --- \\
\hline
MAE & 13 & 7 & 7 & 6 & \\
\hline
\end{tabular}
\label{tab_bulk_modulus}
\end{table}
The calculated and experimental bulk moduli $B_0$ are given in TABLE \ref{tab_bulk_modulus}. PBE generally underestimates the bulk moduli except for the bulk modulus of Li, with a overall MAE of about 13\,GPa. PBE0 matches the experimental results with a MAE of only 7\,GPa very well, and except for $\upbeta$-MnO$_2$ the bulk modulus is typically overestimated. However, this comparison between theory and experiment has to be made with care, as in contrast to formation energies and lattice constants that can be measured quite accurately, the experimental bulk moduli have larger uncertainties. Especially, in the case of $\upbeta$-MnO$_2$ the experimental values of the bulk modulus in the literature show large differences depending on the method which was used.\cite{Haines1995} From our calculations the reported value of 260\,GPa fits the trend that the experimental values are between the PBE and PBE0 results better. The HSE06 results are again very similar to the PBE0 results, promoting the general use of HSE06 instead of PBE0 for these types of systems because the same accuracy is obtained with less computational effort. For the PBE0r results we would expect that the values of the oxide benchmark systems are in between the results of PBE and PBE0, which is indeed confirmed. The PBE0r MAE of 6\,GPa is similar to the ones of PBE0 and HSE06 and thus in very good accordance with the experimental measurements.
\\
All four functionals predict that the bulk modulus increases upon lithiation of Li$_x$Mn$_2$O$_4$ in the range $0\leq x\leq1$. The bulk modulus of tetragonal Li$_2$Mn$_2$O$_4$ is however predicted to be smaller than the one of LiMn$_2$O$_4$ with the exception of the PBE0 bulk modulus of Li$_2$Mn$_2$O$_4$ which is equal to the one of LiMn$_2$O$_4$.

\subsection{Band Gaps}

Important properties in electronic applications and in optical absorption are the indirect and direct band gaps, respectively. The indirect band gap is the smallest overall energetic difference between occupied and unoccupied states. For the direct band gap, the differences between highest occupied and lowest unoccupied state are individually calculated for each $\mathbf{k}$-point, and then the minimum of those is determined. The calculated direct band gaps are given in TABLE \ref{tab_band_gap} and compared to spectroscopic data. In case of molecular water, the first spin-allowed electronic transition is considered.
\begin{table}[H]
\centering
\caption{Calculated direct band gaps or first spin-allowed electronical transitions of the non-metallic benchmark systems in eV compared to experimental data. In the calculation of the MAE the differences to the given experimental data excluding H$_2$O(g) are used including predicted zero band gaps.}
\begin{tabular}{lrrrrr}
\hline
System & PBE & PBE0 & HSE06 & PBE0r & Exp. \\
\hline
H$_2$O(g) & 7.1 & 10.1 & 9.3 & 6.3 & 7.4\cite{Seki1981} \\
Li$_2$O$_2$ & 2.0 & 5.3 & 4.5 & 2.1 & --- \\
Li$_2$O & 5.0 & 7.3 & 6.6 & 5.0 & 6.0\cite{Uchida1980} \\
MnO & 0.1 & 4.9 & 4.1 & 2.3 & 4.1\cite{Kurmaev2008} \\
Mn$_3$O$_4$ & 0.9 & 3.9 & 3.1 & 1.6 & 1.9\cite{Hirai2014} \\
$\upalpha$-Mn$_2$O$_3$ & 0.1 & 3.0 & 2.2 & 0.8 & 1.2\cite{Javed2012} \\
$\upbeta$-MnO$_2$ & 0.3 & 2.3 & 1.6 & 0.4 & 1.0\cite{Sherman2005} \\
LiMnO$_2$ & 0.5 & 4.0 & 3.2 & 1.5 & --- \\
$\uplambda$-Mn$_2$O$_4$ & 1.4 & 4.3 & 3.6 & 2.2 & --- \\
Li$_{0.5}$Mn$_2$O$_4$ & 0.0 & 3.2 & 2.4 & 0.9 & --- \\
LiMn$_2$O$_4$ & 0.0 & 2.7 & 1.9 & 1.1 & 1.2\cite{Ouyang2006} \\
Li$_2$Mn$_2$O$_4$ & 1.4 & 4.5 & 3.8 & 2.2 & --- \\
\hline
MAE & 1.5 & 1.5 & 0.7 & 0.7 & \\
\hline
\end{tabular}
\label{tab_band_gap}
\end{table}
Based on the results of Sec.\ \ref{sec_formation} we have already concluded that the self-interaction error is much larger for PBE than for the hybrid functionals. Strong self-interaction leads to a prediction of a too small or even non-existent band gap,\cite{Cramer2009, Mori-Sanchez2008} since localized electrons become less favored, resulting in an increased metallicity of the system. The data in TABLE \ref{tab_band_gap} confirms this trend. The band gaps calculated by PBE are generally smaller than those calculated by PBE0. The PBE band gaps of all given oxides are underestimated with respect to experiment. Some of the (lithium) manganese oxides are even predicted to be metallic, while the PBE0 functional correctly predicts their non-metallic character. However, the band gaps of all benchmark systems are always overestimated by PBE0 compared to the experimental data. The MAE is 1.5\,eV for the given oxide data. Despite the similar description of energetic and structural properties by PBE0 and HSE06, HSE06 predicts for all of these systems smaller band gaps which are in better agreement with experiment exhibiting a MAE for the oxides of 0.7\,eV only.
\\
As PBE underestimates the band gaps while PBE0 yields too large band gaps, there should be an intermediate HF mixing factor which leads to a very good agreement with experiment as in the case of the formation energies (Sec.\ \ref{sec_formation}). We tested different values for the HF mixing factor of Mn in the PBE0r functional and we experienced that increasing the HF mixing factor of Mn opens the band gap of the (lithium) manganese oxides. For example, in the range from 5 to 10{\%} of the HF mixing factor the band gap of LiMn$_2$O$_4$ widens from 0.7 to 1.2\,eV. The chosen intermediate value of 9{\%} does not only lead to a good prediction of the formation energies, but it also yields very accurate band gaps. The MAE for the band gaps of the oxides is only 0.7\,eV, and the individual values are, as expected, in between the PBE and PBE0 gaps. For the covalently bonded H$_2$O this trend is not present. This again confirms that the approximation of the PBE0r functional is not well-suited for covalent bonds, but performs very well for the (lithium) manganese oxides.

\subsection{Density of States}

\begin{figure}[H]
\centering
\includegraphics[width=\columnwidth]{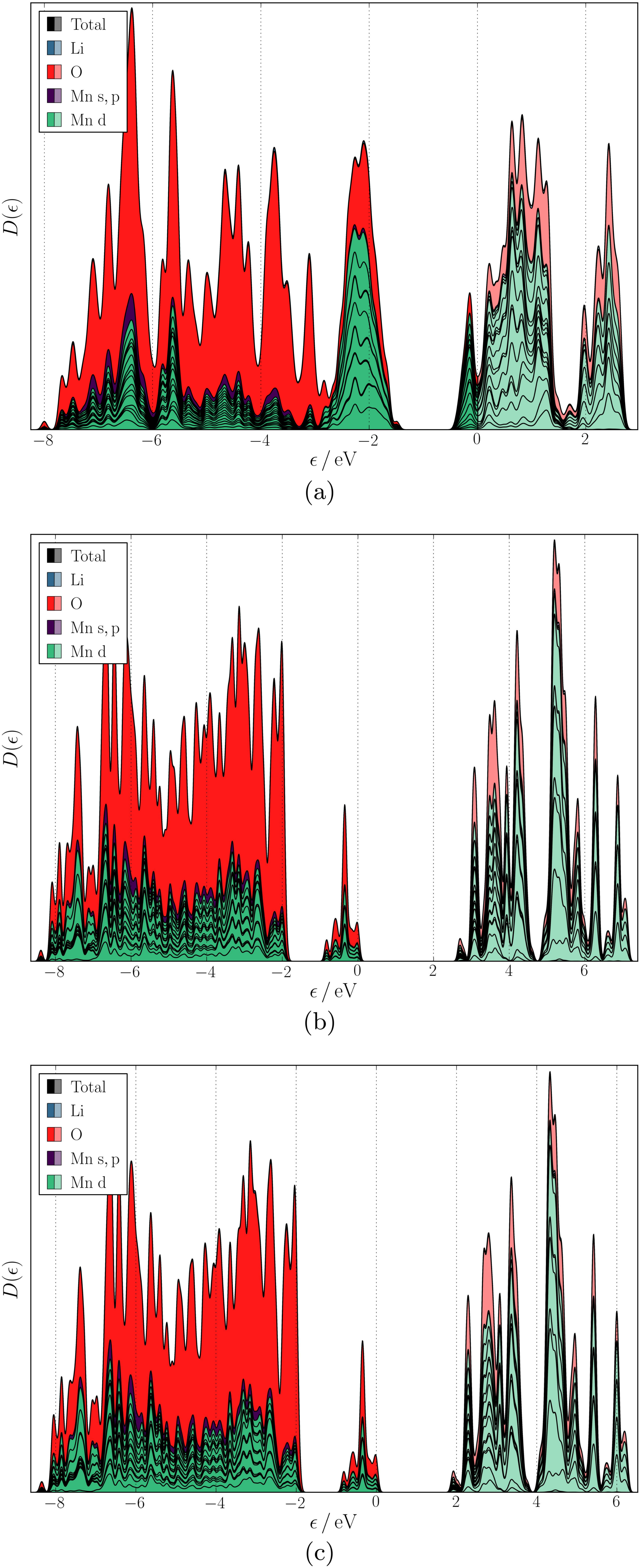}
\caption{The density of states $D(\epsilon)$ of one spin channel plotted for the atoms in the LiMn$_2$O$_4$ unit cell calculated by (a): PBE, (b): PBE0, and (c): HSE06. Unoccupied orbitals are in lighter colors.}\label{fig_dos_FHIaims}
\end{figure}

\begin{figure}[H]
\centering
\includegraphics[width=\columnwidth]{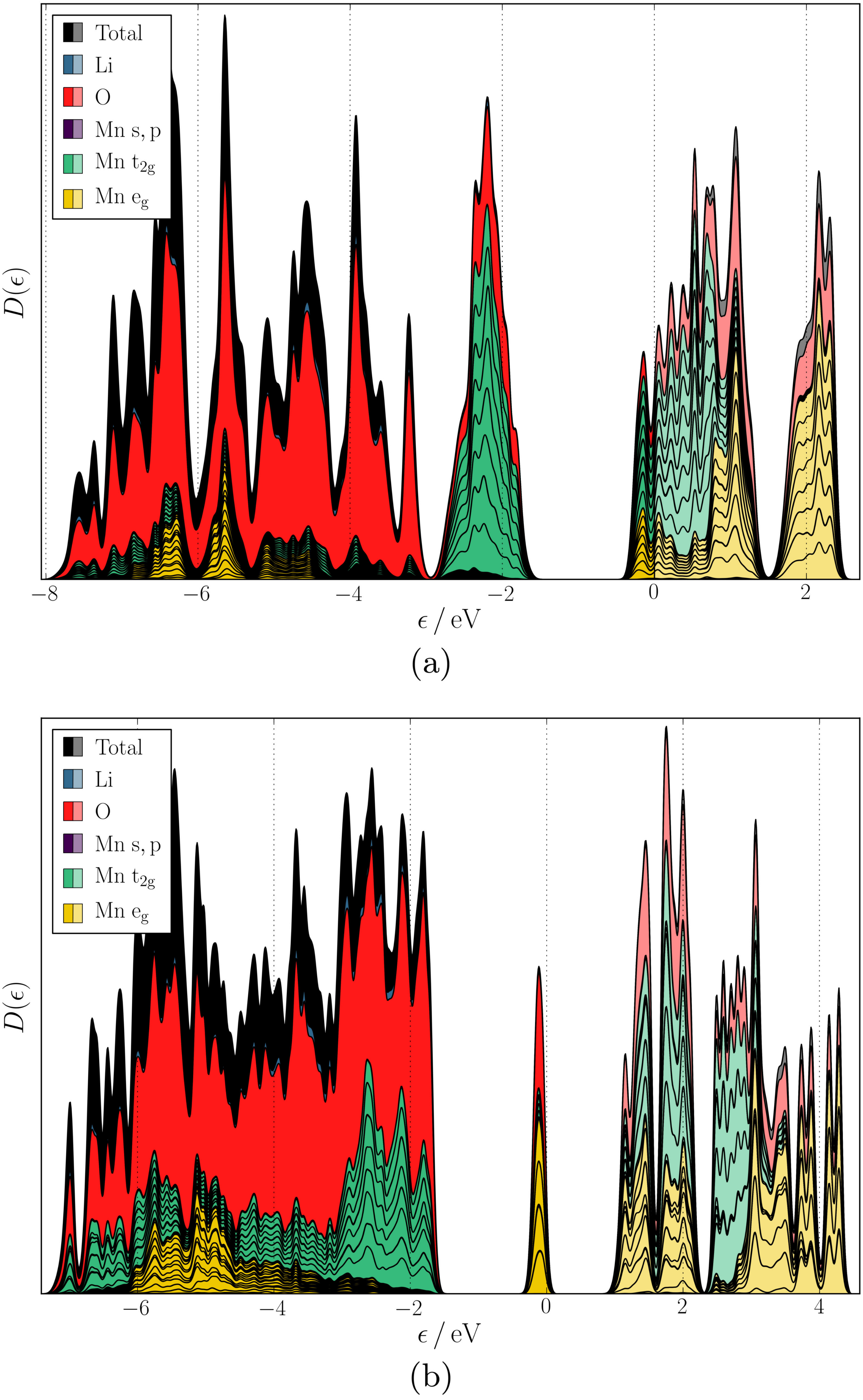}
\caption{The density of states $D(\epsilon)$ of one spin channel plotted for the atoms in the LiMn$_2$O$_4$ unit cell calculated by (a): PBE$_\mathrm{PAW}$ and (b): PBE0r. Unoccupied orbitals are in lighter colors.}\label{fig_dos_CP-PAW}
\end{figure}

\begin{figure}[H]
\centering
\includegraphics[width=\columnwidth]{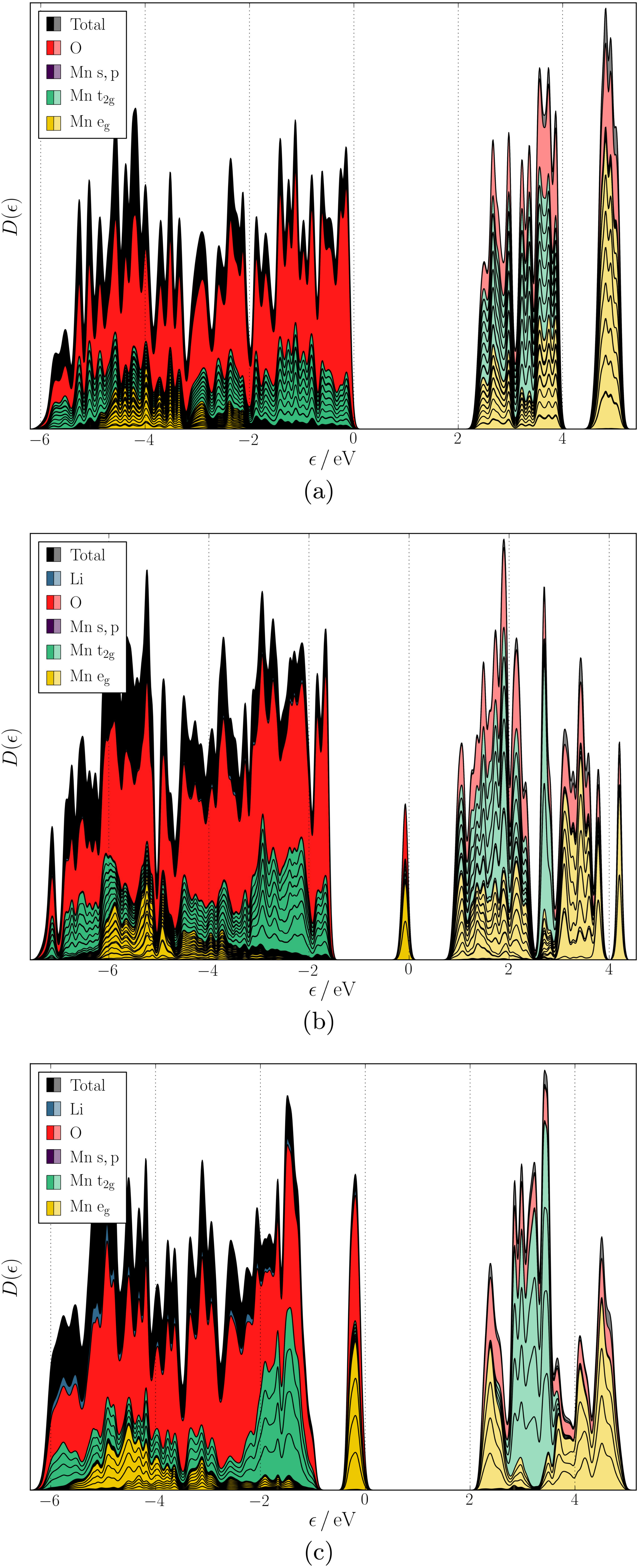}
\caption{The PBE0r density of states $D(\epsilon)$ of one spin channel plotted for the atoms (a): in the $\uplambda$-Mn$_2$O$_4$ unit cell (48 atoms), (b): in the Li$_{0.5}$Mn$_2$O$_4$ unit cell (52 atoms), and (c): in the Li$_2$Mn$_2$O$_4$ unit cell (32 atoms). Unoccupied orbitals are shown in lighter colors.}\label{fig_dos_lixmn2o4}
\end{figure}

The Kohn-Sham density of states (DoS) $D(\epsilon)$ provides detailed insights into the electronic structure of a system, and, for example, the influence of Li insertion can be studied. Furthermore, some features like band gaps or orbital occupations can be directly compared to experimental data.
\\
First, we will compare the DoS of LiMn$_2$O$_4$ calculated by PBE, PBE0, HSE06, and PBE0r (FIG.\ \ref{fig_dos_FHIaims} and \ref{fig_dos_CP-PAW}) in order to understand why the PBE results are less accurate than the results of the hybrid functionals (more information on the calculation procedure of the DoS in both DFT codes are given in the SI). Afterwards, we will study the effect of Li insertion on the electronic structure using the PBE0r DoS of Li$_x$Mn$_2$O$_4$ with $x=0, 0.5, 1,$ and $2$ (FIG.\ \ref{fig_dos_CP-PAW} (b) and \ref{fig_dos_lixmn2o4}). In FIG.\ \ref{fig_dos_FHIaims}, \ref{fig_dos_CP-PAW}, and \ref{fig_dos_lixmn2o4} the peaks of the DoS are broadened for visualization which is done differently by FHI-aims and CP-PAW and must be considered in the comparison. All figures show only the DoS of one spin channel since the DoS of the other spin channel is identical for these optimized structures. The partial DoS of Li, O, Mn s and p, and Mn d orbitals are shown as a stacked plot in front of the total DoS plot. For Mn d states the individual contribution of each atom in the unit cell is plotted separately. In the DoS plots obtained from PBE$_\mathrm{PAW}$ and PBE0r calculations, the Mn d partial DoS is further splitted into the t$_\mathrm{2g}$ and e$_\mathrm{g}$ contributions.
\\
The PBE DoS calculated by FHI-aims (FIG.\ \ref{fig_dos_FHIaims} (a)) and CP-PAW (FIG.\ \ref{fig_dos_CP-PAW} (a)) are very similar. They reveal the missing band gap in those calculations. Moreover, they show that all Mn atoms have the same electron density corresponding to an oxidation state of 3.5. Accordingly, the d electrons are not localized at the Mn ions, because then the oxidation state would be an integer number. Instead, they are strongly delocalized electrons. CP-PAW enables resolving the d electron density into the t$_\mathrm{2g}$ and e$_\mathrm{g}$ contributions, which are present in an octahedral ligand field. This reveals that all Mn d electrons have mainly t$_\mathrm{2g}$ character. The stability underestimation of hs-states is again typical for self-interaction errors.\cite{Cramer2009}
In conclusion, PBE predicts metallic LiMn$_2$O$_4$ with all Mn being in the low-spin Mn$^{3.5}$ state. The MnO$_6$ octahedra are slightly distorted. The PBE$_\mathrm{PAW}$ result shows that two Mn-O bonds are 1.94\,{\AA} and four are 1.97\,{\AA}. However, if this bond length difference is compared to the one of the JT distorted Mn$^\text{III}$ in case of the hybrid functionals mentioned later, it is obvious that this is not an accurate description of the JT distortion. The PBE outcome is contradicting experimental results which show that LiMn$_2$O$_4$ has a band gap and one half of the Mn is in the JT distorted hs-Mn$^\text{III}$ state while the other half is in the Mn$^\text{IV}$ state.
\\
The DoS of the hybrid functionals (FIG.\ \ref{fig_dos_FHIaims} (b) and (c) and FIG.\ \ref{fig_dos_CP-PAW} (b)) are very similar but clearly differ from the PBE DoS. The main effect of exact exchange is the self-interaction correction, which shifts filled electron levels downward. For manganese oxides this shifts the majority spin t$_\mathrm{2g}$ states into the O 2p valence band. Similarly, it shifts the majority spin e$_\mathrm{g}$ orbitals below the minority spin t$_\mathrm{2g}$ orbitals and thus favors high-spin Mn.
\\
In particular PBE0 and HSE06 yield an almost equal DoS except for the size of the band gap. This is in accordance with the similar energetic and structural results. Hence, the reliability of the range-separated hybrid functional is again confirmed. The Mn d orbitals are split into two states of different energy due to the ligand field. In the PBE0r DoS this separation into the t$_\mathrm{2g}$ and e$_\mathrm{g}$ states is highlighted. The lower band is similar for each Mn since both Mn$^\text{III}$ and Mn$^\text{IV}$ have three t$_\mathrm{2g}$ electrons. However, FIG.\ \ref{fig_dos_CP-PAW} (b) shows that only half of the Mn in one spin channel in the LiMn$_2$O$_4$ unit cell (four of eight) have an e$_\mathrm{g}$ electron which corresponds to the energetically highest occupied band. Thus, there are two different types of Mn in the LiMn$_2$O$_4$ unit cell: eight Mn$^\text{III}$ and eight Mn$^\text{IV}$ (in both cases four with spin up and four with spin down). Separately plotted DoS for both Mn states are given in the SI. They show that the t$_\mathrm{2g}$ and e$_\mathrm{g}$ electrons of a Mn have always the same spin direction, i.e.\ the Mn is in the hs-state. The hs-Mn$^\text{III}$ leads to a JT distortion of the corresponding MnO$_6$ octahedron.
\\
A geometrical study of the structure which was optimized by PBE0 or HSE06 using the experimental lattice parameters, i.e.\ a cubic unit cell, shows that two Mn-O bonds are 1.92\,{\AA}, two are 2.05\,{\AA}, and two are 2.07\,{\AA} for the MnO$_6$ octahedra of Mn$^\text{III}$. PBE0r predicts the averaged bond distances 1.93\,{\AA}, 2.05\,{\AA}, and 2.10\,{\AA} for these bonds. These confirm the presence of a JT distortion. The MnO$_6$ octahedra of the Mn$^\text{IV}$ are not JT distorted. PBE0 and HSE06 predict two Mn-O bond lengths of 1.89\,{\AA} and four of 1.90\,{\AA}, i.e.\ the MnO$_6$ octahedron is only very slightly distorted for Mn$^\text{IV}$. PBE0r yields for these bond lengths the averaged values 1.89\,{\AA} and 1.94\,{\AA}. The Mn-O bond lengths also show that the MnO$_6$ octahedra of Mn$^\text{III}$ are larger than those of Mn$^\text{IV}$ because of the additional e$_\mathrm{g}$ electron of Mn$^\text{III}$ in the antibonding Mn-O orbital. All these features of the DoS of the hybrid functionals are in accordance with experimental results.
\\
The Li$_x$Mn$_2$O$_4$ structure does not undergo any structural transformations if the Li content is varied in the range $0\leq x\leq1$. Only the lattice constant and the number of JT distorted MnO$_6$ octahedra decrease with decreasing Li content. Thus, one would expect that the DoS changes only slightly except for the number of e$_\mathrm{g}$ electrons. This is confirmed by the data shown in FIG.\ \ref{fig_dos_CP-PAW} (b) and \ref{fig_dos_lixmn2o4} (a) and (b). The number of e$_\mathrm{g}$ electrons is in all cases equal to the number of Li$^+$ ions. There is no Mn$^\text{III}$ present in $\uplambda$-Mn$_2$O$_4$. As a consequence, the band gap of $\uplambda$-Mn$_2$O$_4$ is much larger than the band gaps of Li$_x$Mn$_2$O$_4$ with $0<x\leq1$: the band gap in $\uplambda$-Mn$_2$O$_4$ is between the mixed O 2p/Mn t$_\mathrm{2g}$ valence band and the unoccupied d states, while in Li$_x$Mn$_2$O$_4$ the highest occupied orbitals are the mixed O 2p/Mn e$_\mathrm{g}$ states which are higher in energy as the O 2p/Mn t$_\mathrm{2g}$ states. The band gap between the O 2p/Mn e$_\mathrm{g}$ states and the unoccupied d states is determined by the much smaller JT splitting of the e$_\mathrm{g}$ states.
The DoS of Li$_2$Mn$_2$O$_4$ (FIG.\ \ref{fig_dos_lixmn2o4} (c)) differs more from the DoS of Li$_x$Mn$_2$O$_4$ with $0\leq x\leq1$ because the unit cell is tetragonally distorted. However, the trend of the increasing number of e$_\mathrm{g}$ electrons is retained. All Mn in the unit cell have one e$_\mathrm{g}$ electron which means that they are all in the Mn$^\text{III}$ state. The JT distortion opens a gap between both e$_\mathrm{g}$ states of the same spin channel. These two states are the highest occupied and lowest unoccupied state for Mn$^\text{III}$. This is shown in the DoS of one Mn$^\text{III}$ which is given in the SI. In Li$_2$Mn$_2$O$_4$ only Mn$^\text{III}$ ions are present and the entire structure is tetragonally distorted. Therefore, the band gap is larger than for Li$_x$Mn$_2$O$_4$ with $0<x\leq1$ because only strongly JT distorted Mn$^\text{III}$ are present. Moreover, the JT distortion decreases the difference between the t$_\mathrm{2g}$ and e$_\mathrm{g}$ state. This trend is also observed if we compare FIG.\ \ref{fig_dos_CP-PAW} (b) with FIG.\ \ref{fig_dos_lixmn2o4} (c).
\\
The DoS of the various (lithium) manganese oxides can be analyzed in a similar way. The hs-Mn state is preferred for all given systems by the hybrid functionals. The PBE0r spin values of the Mn, from which one can derive their oxidation states, are given in the SI.

\section{Conclusion}

The PBE, PBE0, HSE06, and PBE0r exchange-correlation functionals have been benchmarked for various Li$_x$Mn$_y$O$_z$ systems. The deviations of predicted energetic, structural, and electronic properties from experimental data can in general be reduced by the inclusion of exact exchange contributions. Neglecting the long-range exact exchange terms is proven to be a very good approximation. The quality of the HSE06 results is very similar to that of the PBE0 results or even better in the case of band gaps. However, the average computation time per self-consistency cycle of the benchmark systems is reduced by a factor of 0.8 which makes the HSE06 functional preferable to the PBE0 functional for the given types of systems. The experimental formation energies, intercalation potentials, bulk moduli, and band gaps of the studied (lithium) manganese oxides are in between the PBE and PBE0 results. Consequently, an intermediate HF mixing factor as used in the PBE0r functional can further decrease the deviation in many cases. In PBE0r an individual HF mixing factor is assigned to each element. They were determined in a systematic search on a grid of values minimizing the differences to given reference data. Due to the restriction of using only on-site HF exchange terms in PBE0r its average computation time per iteration of the benchmark systems is only increased by a factor of 1.3 compared to PBE$_\mathrm{PAW}$, which is substantially less expensive than the PBE0 and HSE06 functionals. However, this approximation is not well-suited for strongly covalently bonded or metallic systems while it provides reliable results for highly correlated systems. The admixture of on-site HF exchange terms yields most of the improvement from GGA to hybrid functionals in systems with narrow, partially filled d and f shells as they are present in many transition metal oxides. Consequently, the accuracy of the results for the (lithium) manganese oxides is greatly improved compared to PBE. We conclude that PBE0r has the optimum cost-benefit ratio for these types of systems. In summary, the hybrid functionals PBE0, HSE06, and PBE0r agree well with experiment.
\\
Theoretical calculations for the lithium manganese oxide spinel Li$_x$Mn$_2$O$_4$, with $0\leq x\leq2$, agree well with experimental results if hybrid functionals are used. The calculated formation energies, the trend of the intercalation potentials, the equilibrium lattice constants, the bulk moduli, and the band gaps are in accordance with experimental data. Several essentially degenerate antiferromagnetic states exist close to the energetically lowest configuration of Li$_x$Mn$_2$O$_4$. A formation of two oxidation states, +III and +IV, is observed for Mn. The Mn$^\text{III}$ ions are in the hs-state and the corresponding MnO$_6$ octahedra are JT distorted. 

\begin{acknowledgments}
We thank the DFG (SFB 1073 project B03 and C03, project 217133147, project 405832858) for financial support. We gratefully acknowledge the funding of this project by computing time provided by the Paderborn Center for Parallel Computing (PC$^2$). JB is grateful for a DFG Heisenberg professorship (BE3264/11-2, project 329898176).
\end{acknowledgments}

\bibliography{bibliography}

\end{document}